\documentclass[twocolumn,showpacs,showkeys,amsmath,amssymb]{revtex4}

\usepackage{graphicx}

\begin{document}

\title{Strong deflection limit of black hole gravitational lensing
with arbitrary source distances}

\author{V. Bozza$^{a,b}$, G. Scarpetta$^{a,b,c}$}

\affiliation{$^a$ Dipartimento di Fisica ``E.R. Caianiello'',
Universit\`a di Salerno, via Allende,
I-84081 Baronissi (SA), Italy.\\
  $^b$  Istituto Nazionale di Fisica Nucleare, Sezione di
 Napoli. \\
 $^c$ International Institute for Advanced Scientific Studies, Vietri sul Mare (SA), Italy}

\date{\today}

\begin{abstract}
The gravitational field of supermassive black holes is able to
strongly bend light rays emitted by nearby sources. When the
deflection angle exceeds $\pi$, gravitational lensing can be
analytically approximated by the so-called strong deflection
limit. In this paper we remove the conventional assumption of
sources very far from the black hole, considering the distance of
the source as an additional parameter in the lensing problem to be
treated exactly. We find expressions for critical curves, caustics
and all lensing observables valid for any position of the source
up to the horizon. After analyzing the spherically symmetric case
we focus on the Kerr black hole, for which we present an
analytical 3-dimensional description of the higher order caustic
tubes.
\end{abstract}



\maketitle

\section{Introduction}

Black holes have always attracted the interest of all physicists
who hope to see General Relativity at work in a completely
non-perturbative regime, outside any Post-Newtonian expansion.
Since most of the information we receive from black holes and
their surroundings is in the form of electromagnetic waves, one of
the fundamental problems to be faced is the propagation of such
waves in a black hole spacetime. The situation can be possibly
complicated by an accretion disk formed by neutral plasma
\cite{BroBla}. For wavelengths at which observations are typically
lead, the geometrical optics approximation provides a very robust
description for the propagation of the light rays, defined as the
lines orthogonal to the wavefronts. In all situations in which the
plasma physics has little effect on the rays trajectories, the
light rays simply follow null geodesics. Then all questions that
involve the propagation of an electromagnetic signal require
integration of the null geodesics equation. In the case of Kerr
metric, the null geodesics have been expressed by Carter in terms
of first integrals through the separation of the Hamilton-Jacobi
equation \cite{Car}. Then, the integrals involved in these
geodesics can be solved in terms of elliptic functions (see e.g.
\cite{RauBla,CFC}). Although these analytical solutions are not
particularly illuminating by themselves, they can be successfully
employed to build fast and accurate numerical codes
\cite{Vie,RauBla,CFC,BecDon}, by which one can get particular
answers, such as the shape of the iron K-line or the appearance of
the accretion flow into the black hole in a future hypothetical
high-resolution image \cite{Lum,Accret,BroLoe}.

The problem of finding the null geodesics connecting a source to
an observer in a curved background is usually referred to as
gravitational lensing. It has been pointed out by many authors
that in a black hole spacetime there are infinitely many possible
trajectories for the photons emitted by a point-source to reach
the observer \cite{Dar,Lum,Oha,Nem,VirEll,BCIS,HasPer}. For each
of these trajectories the observer will detect an independent
image of the original source. The images can be classified
according to the number of loops performed by the corresponding
photons around the black hole. One starts from the primary and
secondary image, which are formed by photons performing no loops.
These are already present in the classical weak deflection limit
of gravitational lensing. Besides these, there are two infinite
sequences of higher order images, formed by photons performing one
or more loops around the black hole before reaching the observer.
These images are progressively fainter and appear closer and
closer to the apparent shadow cast by the black hole on the sky.

The higher order images contribute much less than the primary and
secondary images to the total flux and are often completely masked
in situations in which they are not separable from the main images
\cite{Pet}. Therefore, the best chances to observe higher order
images are offered by a black hole as massive and close to us as
possible, so that its apparent angular size is the largest. The
natural candidate is the black hole at the center of our Galaxy,
identified with the radio-source Sgr A*. This is believed to be a
$3.6\times 10^6 M_\odot$ supermassive black hole slowly accreting
material from the surrounding environment \cite{MelFal,Eis}. Its
distance is $D_{OL}=8$kpc, so that the Schwarzschild radius
$R_{Sch}$ of this black hole spans an angle of roughly 9 $\mu$as
in our sky. A resolution of this order of magnitude is needed to
detect the higher order images, besides the requirement of
negligible absorption in the wavelength of the emitted signal by
the surrounding material. In spite of these difficulties, the
detection of higher order images of sources around Sgr A* should
be at hand of future interferometers operating in the sub-mm
range, where one expects to detect higher order images of bright
spots on the accretion disk \cite{BroLoe}, or in the X-ray band,
where Low Mass X-Ray Binaries and other sources are active
\cite{Muno,RauBla,KerObs,KerGen}. Such images would be invaluable
witnesses of the strong gravitational field around the
supermassive black hole at the center of our Galaxy; their direct
observation would thus be of striking importance in the
confirmation of our gravitational theory.

In addition to the considerations about their importance, higher
order images can boast two advantages with respect to lower order
ones: if the black hole has non-negligible spin, they can easily
form large arcs and additional images because their caustics have
a larger angular extension compared to the first caustic;
secondly, they enjoy a relatively much simpler analytical
description.

The treatment of higher order images can take advantage of the
fact that the deflection diverges logarithmically when the impact
parameter reaches a minimum value. Then, the higher order images
can be obtained by a simplified lens equation where the deflection
terms are replaced by the first terms of their expansions in terms
of the impact parameter. This procedure is analogous to the weak
deflection limit but sets its starting point in the opposite
regime and is thus conventionally called {\it strong deflection
limit}. It was firstly used by Darwin in 1959 for the
Schwarzschild black hole \cite{Dar} and then revived several times
\cite{Lum,Oha,Nem,BCIS} until it was generalized to all
spherically symmetric black holes \cite{Boz1}. This method was
then applied to several interesting black hole metrics, also
motivated by string and alternative theories
\cite{Spheric,BorInf,BHdS}. The time delay calculation for higher
order images was done in Ref. \cite{TimDel}. The method was
recently extended to the presence of external shear fields in a
setup analogous to the Chang \& Refsdal lens \cite{BozMau}. The
extension to Kerr metric has required several steps, starting from
the purely equatorial case \cite{BozEq} to the case of generic
trajectories with equatorial observers \cite{KerObs}, and finally
to the general case \cite{KerGen} (in the latter two works, the
treatment is however limited to the second order in the black hole
spin). An application to the Kerr-Sen dilaton-axion black hole has
also been performed \cite{KerSen}. Recently Iyer and Petters have
found an alternative expansion parameter that significantly
reduces the discrepancy between the strong deflection limit and
the exact deflection formula \cite{IyePet}. They have also
explored the next to leading order terms in the strong deflection
expansion.

The strong deflection limit allows a simple analytic investigation
of the gravitational lensing properties of any black hole metric
in a well-defined limit, in which the results are easily
comparable from one metric to another. For the Kerr metric it has
been able to provide the first analytical formulae for the
caustics and the critical curves involving higher order images and
for lensing of sources near caustics. However, in the formulation
used up to now, it has just been applied to sources very far from
the black hole, so that the limit $D_{LS} \gg R_{Sch}$ has been
taken in all equations (here $D_{LS}$ is the distance of the
source from the black hole and $R_{Sch}=2G M/c^2$ is the
Schwarzschild radius of the black hole).

The purpose of this paper is to remove the limitation to very far
sources, enlarging the investigation of gravitational lensing in
the strong deflection limit to sources placed at arbitrary
distances from the black hole. We will thus be able to discuss the
mathematical structure of the lensing problem and all the lensing
observables taking $D_{LS}$ as an additional parameter. We will
show that the strong deflection limit is well-defined even for
sources inside the photon sphere, so that our discussion can be
safely pushed up to sources lying just outside the horizon of the
black hole. Similarly, in the Kerr metric we will be able to
describe the caustic hypersurface from infinite radial distance up
to the horizon.

This paper is structured as follows. Sect. \ref{Sec Spherical}
contains the new outline of the strong deflection limit for
spherically symmetric black holes with arbitrary source position;
it analyzes the lens equation and observables and discusses the
Schwarzschild metric as a simple example. Sect. \ref{Sec Kerr}
contains the extension of Kerr black hole lensing to arbitrary
source distances, with some details moved to the appendix. Section
\ref{Sec Conclusions} contains the conclusions.

\section{Spherically symmetric black holes} \label{Sec Spherical}

In this section we shall present an updated version of the method
outlined in Ref. \cite{Boz1}. Besides including the finiteness of
source and observer distances from the black hole, we also make
some more slight refinements that allow further generalization of
the method and more physical insight. We stress the importance of
the study of spherically symmetric black holes as propaedeutic to
the investigation of the Kerr metric, to be tackled in the next
section.

As in Ref. \cite{Boz1}, we start from three basic assumptions on
the spacetime metric:

a) The spacetime is stationary and spherically symmetric, so that
the line element can be written in the form
\begin{equation}
ds^2=A(r)dt^2-B(r)dr^2-C(r)\left( d\vartheta^2+\sin^2\vartheta
d\phi^2 \right). \label{Metric}
\end{equation}

b) We assume that our metric is asymptotically flat, so that for
$r\rightarrow \infty$ we have
\begin{equation}
A(r)\rightarrow 1, ~~~~B(r)\rightarrow 1, ~~~~C(r)\rightarrow r^2.
\end{equation}

c) Furthermore, we assume that the function $C(r)/A(r)$ has one
minimum at $r_m>0$, corresponding to the radius of the photon
sphere $r_m$ \cite{CVE}.

In some gravitational theories, the photons do not follow
geodesics of the background geometry, but the self-interaction
makes them follow geodesics with respect to some effective metric
\cite{BorInf}. These particular cases can fit into our treatment,
provided that one uses the effective metric felt by the photons.
Assumption (b) can be generalized to spacetimes that are
asymptotically conformal to flat, thanks to the conformal
invariance of null geodesics. In this way one can include e.g.
black holes with a cosmological constant \cite{BHdS}.

\begin{figure}
\resizebox{\hsize}{!}{\includegraphics{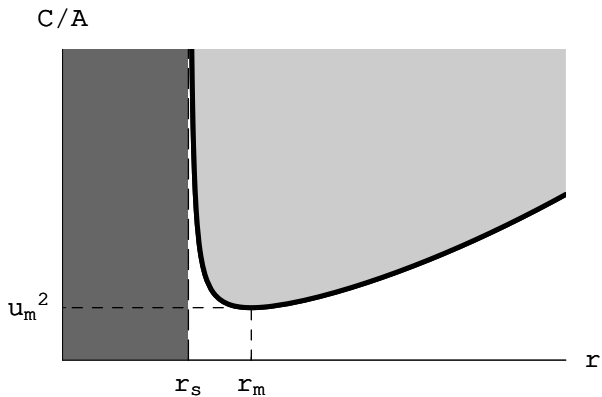}}
\caption{The function $C(r)/A(r)$ plotted for the Schwarzschild
metric ($A=1-R_{Sch}/r$, $C=r^2$). In this case the static limit
is $r_s=R_{Sch}$ and the photon sphere radius is $r_m=1.5
R_{Sch}$. The light-gray region is forbidden because it
corresponds to imaginary $\dot r$. The dark-gray region lies below
the horizon of the black hole and is uninteresting for
gravitational lensing.}
 \label{Fig C/A}
\end{figure}

Let us give a closer look at assumption (c) and specifically at
the function $C/A$. We can note that asymptotic flatness drives
the function $C/A$ to approach $r^2$ at very large $r$. Moreover,
in any metric admitting a static limit $r_s$, such that
$A(r_s)=0$, the function $C/A$ diverges at $r_s$. So, by
continuity it must have at least one minimum greater than $r_s$.
As a useful reference, in Fig. \ref{Fig C/A}, we plot the function
$C/A$ for the Schwarschild spacetime, where it assumes the form
$r^3/(r-R_{Sch})$. Following Ref. \cite{CVE}, all stationary
points of $C/A$ are technically photon spheres. However, maxima
are not significant for gravitational lensing, since they are only
accessible to locally emitted photons. We are actually interested
in minima of $C/A$, since they give rise to logarithmic
divergences in the deflection angle, as it will be clear later.
Although it is relatively easy to find metrics which also develop
a maximum in $C/A$ (e.g. Reissner-Nordstr\"{o}m with superextremal
charge), it seems difficult to imagine a metric developing a
second minimum. Such a metric would require a quite exotic source
to be sustained as a viable solution of gravitational equations.
Through assumption (c), we are discarding these too problematic
spacetimes and stick to more reasonable metrics with only one
minimum for $C/A$.

Now let us start the calculation of the photon deflection. The
spherical symmetry allows us to choose the equatorial plane as the
plane where the entire motion of the photon takes place, so that
$\vartheta=\pi/2$ and $\dot \vartheta=0$, where the dot denotes
derivative with respect to the affine parameter.

The dynamics of the photon can be derived from the Lagrangian (see
e.g. \cite{Cha} for a complete discussion of null geodesics in
Schwarzschild and Kerr spacetimes)
\begin{equation}
\mathcal{L}=-\frac{1}{2}g_{\mu\nu}\dot x^\mu\dot x^\nu.
\end{equation}
The coordinates $t$ and $\phi$ are cyclic so that their conjugate
momenta are constants of motion. They can be identified with the
specific energy $E$ and angular momentum $J$
\begin{eqnarray}
&&  E=A(r)\dot t \label{tdot}\\
&& J=C(r)\dot \phi. \label{phidot}
\end{eqnarray}
We can choose the orientation of the polar axis so that $J>0$.

The last constant of motion comes from the fact that the photon
moves along null geodesics of the metric (\ref{Metric}) so that
$g_{\mu\nu}\dot x^\mu \dot x ^\nu=0$. From this equation, we
derive the expression of $\dot r$:
\begin{equation}
\dot r=\pm \frac{E}{\sqrt{B C}}\sqrt{\frac{C}{A}-\frac{J^2}{E^2}}.
\label{rdot}
\end{equation}

The angle formed by the spatial components of the photon momentum
$p^i=(\dot r, \dot \phi)$ with a normalized vector tangent to a
sphere centered on the black hole $t^i=(0,1/\sqrt{C})$ is
\begin{equation}
\vartheta=\arccos \frac{-g_{ij}t^ip^j}{|p||t|}=\arccos\left(
\frac{J}{E} \sqrt{\frac{A}{C}} \right). \label{EleAng}
\end{equation}

So, in any point of the photon trajectory, the knowledge of the
value of the combination
\begin{equation}
u=\frac{J}{E},
\end{equation}
allows us to calculate the angles formed by the photon with
respect to the radial and tangent directions. The photon moves
radially if $u=0$ and tangentially in points such that
$u=\sqrt{C/A}$. It is also easy to prove that for those photons
reaching the asymptotic flat region, this quantity equals the
impact parameter, defined as the distance between the black hole
and the asymptotic trajectory followed by the photon. On the basis
of its immediate connection with the observed direction of the
photon, we will eliminate $J$ in favor of $u$, where possible.

Inversion of the radial motion can occur only at the points that
make the argument of the square root vanish in Eq. (\ref{rdot}),
which correspond to points of instantaneous tangential motion by
virtue of Eq. (\ref{EleAng}). However, assumption (c) states that
the function $C/A$ has a single minimum at $r_m$. So, a quick look
at Fig. \ref{Fig C/A} convinces that the equation
\begin{equation}
\frac{C(r_0)}{A(r_0)}=u^2 \label{ur0},
\end{equation}
has real roots only if $u>u_m$, with
\begin{equation}
u_m=\sqrt{\frac{C_m}{A_m}},\label{um}
\end{equation}
where we have introduced the short notation $A_m\equiv A(r_m)$ and
similarly for $B$ and $C$.

At this point it is convenient to distinguish the case
$D_{LS}>r_m$ (source outside the photon sphere) from the case
$D_{LS}<r_m$ (source inside the photon sphere). We shall finally
find that they both lead to the same expression for the deflection
of a photon in the strong deflection limit, given by Eq.
(\ref{alphaSDL}). The next two subsections deal with the details
of the calculations in the two mentioned cases.

\subsection{Source outside the photon sphere}

Let us analyze all possibilities for the radial motion of a photon
emitted by a source outside the photon sphere.

Some photons will leave the source with positive $\dot r$. Since
the photons are emitted at $D_{LS}>r_m$, these photons never meet
inversion points and run towards the asymptotic region without
experiencing any effective deflection by the black hole. If
$D_{OL}>D_{LS}$ some of them reach the observer and give rise to
the primary image, which is not the subject of our analysis
anyway.

Some other photons leave the source with negative $\dot r$. If
$D_{OL}<D_{LS}$, there is still the possibility that some of them
can reach the observer without inverting their motion and form the
primary image. But some other photons do not meet the observer and
run towards the black hole. The photons with $u<u_m$ will
inexorably sink into the black hole, since Eq. (\ref{ur0}) will
admit no real roots. If $u>u_m$, the photons invert their motion
at the largest root of Eq. (\ref{ur0}), which we indicate by $r_0$
and identify with the closest approach distance. After the
inversion in the radial motion, these photons go back towards the
asymptotic region and eventually reach the observer, giving rise
to the secondary and higher order images.

All these considerations can be summarized by saying that light
rays shot at too small impact parameters are swallowed by the
black hole, whereas those shot at larger impact parameters are
just deflected, the limiting value of $u$ being $u_m$. Our
objective is to quantify the deflection of these photons as a
function of $u$.

The azimuthal shift of the photon is
\begin{equation}
\Delta \phi= \int\limits_{\phi_i}^{\phi_f}d\phi=
\int\limits_{D_{LS}}^{r_0}\frac{\dot\phi}{\dot
r}dr+\int\limits^{D_{OL}}_{r_0}\frac{\dot\phi}{\dot r}dr,
\end{equation}
where we have separated the motion of the photon into approach
phase (with $r$ running from $D_{LS}$ to the inversion point
$r_0$) and departure phase (with $r$ running from $r_0$ to
$D_{OL}$). In the first integral we use the expression for $\dot
r$ with the minus sign, in the second integral we use the
expression with the plus sign. Using Eqs. (\ref{phidot}) and
(\ref{rdot}), and reversing the extrema in the first integral, we
have
\begin{eqnarray}
&& \Delta \phi= \Delta \phi_S+\Delta \phi_O \label{Dphi}\\
&&  \Delta\phi_i=\int\limits^{D_{Li}}_{r_0}
u\sqrt{\frac{B}{C}}\left(\frac{C}{A}-u^2\right)^{-1/2}dr,
\label{Dphii0}
\end{eqnarray}
with the short notation $i=O,S$ and $D_{Li}=D_{OL},D_{LS}$.

Note that the integrand diverges at $r_0$, which is defined as the
largest root of the last factor. In order to study the character
of the divergence, it is opportune to perform a detailed analysis
of the function
\begin{equation}
R(r,u)=\frac{C(r)}{A(r)}-u^2, \label{Rru}
\end{equation}
which governs the divergence of the integrand in the lower
extremum. From the previous discussion, we know that $R(r,u)$ has
a minimum at $r_m$ for any fixed value of $u$; it vanishes at $(r_m,u_m)$
by the definition of $u_m$; it vanishes at
$(r_0,u)$ by the definition of $r_0$. These properties can be formalized by
the equations
\begin{eqnarray}
&& \frac{\partial R}{\partial r}(r_m,u_m)=0 \label{DRrmum}\\
&& R(r_m,u_m)=0 \label{Rrmum} \\
&& R(r_0,u)=0 \label{Rr0u}.
\end{eqnarray}

Since we are interested into those trajectories whose inversion
point is very close to the minimum $r_m$, we define a parameter
$\delta$ by the equation
\begin{equation}
r_0=r_m(1+\delta). \label{r0delta}
\end{equation}
Correspondingly, also the impact parameter must be very close to
the minimum. We thus define the parameter $\epsilon$ by the
equation
\begin{equation}
u=u_m(1+\epsilon). \label{ueps}
\end{equation}

Then, inserting (\ref{r0delta}) and (\ref{ueps}) into Eq.
(\ref{Rr0u}) and expanding to the lowest order in $\delta$ and
$\epsilon$, we have
\begin{eqnarray}
&&0=R(r_m,u_m)+ \frac{\partial R}{\partial r}
(r_m,u_m) r_m \delta \nonumber \\
&&+ \frac{1}{2}\frac{\partial^2 R}{\partial r^2} (r_m,u_m) r_m^2
\delta^2 +\frac{\partial R}{\partial u} (r_m,u_m) u_m \epsilon.
\label{epsdelta}
\end{eqnarray}
The first two terms vanish because of Eqs. (\ref{DRrmum}) and
(\ref{Rrmum}) and we are left with a simple relation between
$\delta^2$ and $\epsilon$, which tells us how much the inversion
point $r_0$ differs from the photon sphere radius $r_m$, when we
increase the impact parameter of the photon from the minimum value
$u_m$ to $u$. Given the form of $R(r,u)$ it simply reads
\begin{equation}
\epsilon=\frac{\beta_m}{2u_m^2}\delta^2, \label{epsdelta2}
\end{equation}
with
\begin{equation}
\beta_m = \frac{1}{2}\frac{\partial^2 R}{\partial r^2} (r_m,u_m)
r_m^2=\frac{1}{2}r_m^2\frac{C''_mA_m-A''_mC_m}{A_m^2},
\label{betam}
\end{equation}
where the prime denotes derivative with respect to the argument
and the subscript $m$ means that the result must be evaluated at
$r_m$ as usual. Thus we obtain that $\epsilon$ is of the same
order as $\delta^2$.

Let us analyze the behavior of $R(r,u)$ when $r$ is very close to
$r_0$. We introduce the parametrization
\begin{equation}
r=\frac{r_0}{1-\eta} \label{rzout}
\end{equation}
with $0\le\eta<1$, and study $R(r(\eta),u)$ for small values of
$\eta$. Expanding to the second order in $\eta$, $\delta$, and
first order in $\epsilon$, we find
\begin{eqnarray}
&&R(\eta,u)\simeq R(r_m,u_m)+r_m \frac{\partial R}{\partial
r}
(r_m,u_m)  (\delta+\eta+\eta \delta+\eta^2) \nonumber \\
&&+\frac{1}{2}r_m^2 \frac{\partial^2 R}{\partial r^2} (r_m,u_m)
(\delta+\eta)^2 +\frac{\partial R}{\partial u} (r_m,u_m)
u_m \epsilon.
\end{eqnarray}
Again, the first two terms vanish because of Eqs. (\ref{DRrmum})
and (\ref{Rrmum}). Moreover, the $\epsilon$-term cancels with the
remaining $\delta^2$-term because of Eq. (\ref{epsdelta}). The
behavior of $R(\eta,u)$ for small $\eta$ is thus
\begin{equation}
R(\eta,u)= \beta_m (2\delta
\eta+\eta^2)+o(\eta^2).
\end{equation}

Now that we have found the dominant terms in $R(\eta,u)$
for $\eta$ close to $0$, which corresponds to $r$ close to
$r_0$, we can return to the integral in Eq. (\ref{Dphii0}).
Changing the integration variable from $r$ to $\eta$, the
integration ranges become $[0,\eta_i]$, where
$\eta_i=1-r_0/D_{Li}$. Each integral assumes the form
\begin{equation}
\Delta\phi_i=\int\limits_0^{\eta_i}
u\sqrt{\frac{B(\eta)}{C(\eta)}}\left[R(\eta,u)\right]^{-1/2}\frac{r_0}{(1-\eta)^2}
d\eta. \label{Dphii}
\end{equation}
Now we add and subtract a term containing the divergence of the
integrand at small values of $\eta$. We then separate the
integral into two parts:
\begin{eqnarray}
&& \Delta\phi_i= I_D+I_R \label{Dphii2}\\
&& I_D= \int\limits_0^{\eta_i}
\frac{u_m}{\sqrt{\beta_m}}\sqrt{\frac{B_m}{C_m}}\frac{r_m}{\sqrt{2\delta
\eta+\eta^2}}d\eta \label{ID}\\
&& I_R= \int\limits_0^{\eta_i}
\left[u\sqrt{\frac{B(\eta)}{C(\eta)}}\left[R(\eta,u)\right]^{-1/2}\frac{r_0}{(1-\eta)^2}
\right. \nonumber
\\ &&\left.-\frac{u_m}{\sqrt{\beta_m}}\sqrt{\frac{B_m}{C_m}}\frac{r_m}{\sqrt{2\delta
\eta+\eta^2}}\right]d\eta. \label{IR}
\end{eqnarray}

\begin{figure}
\resizebox{\hsize}{!}{\includegraphics{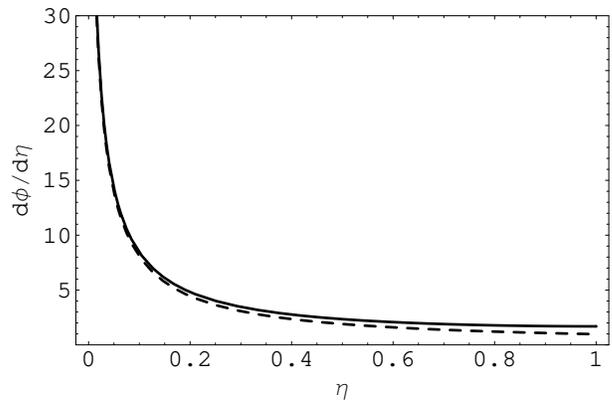}}
\caption{The plot illustrates the integrand of $\Delta\phi_i$ in
Eq. (\ref{Dphii}) as a function of $\eta$ for $\delta=0.025$
($\epsilon=0.001$) in the case of Schwarzschild black hole; the
dashed line is the integrand of $I_D$ in Eq. (\ref{ID}), the
difference between the two curves being the integrand of $I_R$ in
Eq. (\ref{IR}).}
 \label{Fig dphidzout}
\end{figure}

Of course, the sum of the two integrals $I_D$ and $I_R$ is just
the original integral (\ref{Dphii}), but now the first integral is
elementary and reads
\begin{equation}
I_D=r_m\sqrt{\frac{B_m}{A_m\beta_m}}\log
\frac{\eta_i+\delta+\sqrt{\eta_i(\eta_i+2\delta)}}{\delta},
\end{equation}
The divergence for $\delta
\rightarrow 0$    appears   explicitly  in $I_D$, while $I_R$ is the
integral of a regular function and does not diverge any more for
$\delta \rightarrow 0$. Fig. \ref{Fig dphidzout} illustrates an
interesting comparison between the original integrand of Eq.
(\ref{Dphii}) and the integrand of $I_D$ in Eq. (\ref{ID}) taking
the special case of Schwarzschild metric as an example. Although
the integrand of $I_D$ is a drastically simplified form of the
original one which approximates it for $\eta$ very close
to zero, it turns out to be a very good approximation in the whole
range of $\eta$. Only for $\eta$ close to 1 we see
a sensible difference. Such difference is stored in the integrand
of $I_R$. As $\delta$ tends to zero, the contribution by $I_D$
becomes more and more dominant with respect to that of $I_R$ as
the divergence of the integrand becomes stiffer and stiffer. These
considerations are a good premise to the strong deflection limit.

Until now we have done no approximation. We have just added and
subtracted some terms and made some changes of variables. The
expression (\ref{Dphii2}) is still exact. The strong deflection
limit amounts to save the first dominant terms in the expressions
for $I_D$ and $I_R$ as $\delta \rightarrow 0$. We have a
logarithmic divergent term in $I_D$ and then some terms converging
to constant values. We also note that the parametrization
(\ref{rzout}) tends to
\begin{equation}
r=\frac{r_m}{1-\eta}, \label{rzin}
\end{equation}
when $\delta\rightarrow 0$. Consequently, the integration limits
$\eta_i$ tend to $\eta_i=1-r_m/D_{Li}$.

After the truncation of the expansion in $\delta$, we have
\begin{equation}
\Delta \phi_i=a \log \frac{2\eta_i}{\delta} +b_i+o(\delta^0),
\label{Dphii3}
\end{equation}
where the coefficients $a$ and $b_i$ are given by
\begin{eqnarray}
&& a=r_m\sqrt{\frac{B_m}{A_m\beta_m}} \label{aSDL}\\
&& b_i= \int\limits_0^{\eta_i}g_1(\eta)d \eta= \int\limits_0^{\eta_i}Sign(\eta)\,g_1(\eta)d \eta \label{bi}\\
&&g_1(\eta)=
u_m\sqrt{\frac{B(\eta)}{C(\eta)}}\left[R(\eta,u_m)\right]^{-1/2}\frac{r_m}{(1-\eta)^2}
 \nonumber
\\
&&-\frac{u_m}{\sqrt{\beta_m}}\sqrt{\frac{B_m}{C_m}}\frac{r_m}{|\eta|},
\label{g1}
\end{eqnarray}
where $g_1(\eta)$ is just the integrand of $I_R$ in the limit for
$\delta\rightarrow 0$, also implying $u\rightarrow u_m$. The
function $Sign(\eta)$ has been introduced only for uniforming the
expression of $b_i$ to the corresponding terms we shall derive in
section \ref{Sec inside} in the case of a source inside the photon
sphere. For a source outside the photon sphere, the variable
$\eta$ is always positive. Finally, the deflection suffered by the
photon is quantified by the full azimuthal shift $\Delta \phi$,
given by
\begin{equation}
\Delta \phi=a\log\frac{4\eta_O \eta_S}{\delta^2}+b_O+b_S.
\label{alpha1SDL}
\end{equation}

When the source and the observer are very far from the black hole,
it makes sense to define a deflection angle as the difference
between the azimuthal shift suffered by the photon minus $\pi$,
which represents the total azimuthal shift of a photon travelling
in a flat space without the black hole on a rectilinear
trajectory. This concept becomes ill-defined for sources and
observers that are not in the asymptotic flat region.

The fact that source and observer are at finite distances is
encoded in the presence of $\eta_O$ and $\eta_S$. Setting them to
1, the deflection angle so derived coincides with the expression
originally given in Ref. \cite{Boz1}.

With the definition of $\beta_m$ given by Eq. (\ref{betam}) and
Eq. (\ref{Rru}), we can formulate an explicit expression for the
coefficient of the logarithmic term in terms of the metric
functions
\begin{equation}
a=\sqrt{\frac{2B_mA_m}{C''_m A_m-C_mA''_m}}. \label{aSDL2}
\end{equation}

This coefficient is independent of the source and observer
positions.

\subsection{Source inside the photon sphere} \label{Sec inside}

Now let us consider the case in which the source is inside the
photon sphere, but still outside the horizon. The photons leaving
the source with negative $\dot r$ sink into the black hole. As for
the photons leaving with positive $\dot r$, we have two
possibilities: those starting with $u>u_m$ meet an inversion point
before reaching the photon sphere radius $r_m$. Therefore, they
fall back into the black hole. The photons with $u<u_m$ meet no
inversion point and escape towards the asymptotic region. The
observer will thus see the deformed images of a source inside the
photon sphere.

Recalling the relation between $u$ and the angle formed by the
photon momentum with the tangential direction (Eq.
(\ref{EleAng})), we can re-interpret this discussion noting that
only photons shot along the radial direction ($u=0$) or very close
to the radial direction ($u<u_m$) will be able to escape to
infinity. Photons shot at larger angles with respect to the radial
direction invert their motion before crossing the photon sphere.
It is interesting to note that the angle formed with the tangent
direction by photons emitted at $D_{\mathrm{LS}}<r_m$ decreases
until they cross the photon sphere. After that moment, they align
more and more with the radial direction as they move farther and
farther from the black hole.

At first sight, one may think that this situation is very
different from the one described in the previous subsection.
Actually, even in this case it is possible to define a strong
deflection limit, corresponding to photons with $u$ just slightly
smaller than $u_m$. Let us see this in detail.

The azimuthal shift of the photon is
\begin{eqnarray}
&\Delta \phi&= \int\limits_{\phi_i}^{\phi_f}d\phi=
\int\limits_{D_{LS}}^{D_{OL}}\frac{\dot\phi}{\dot r}dr \nonumber
\\ && \int\limits_{D_{LS}}^{D_{OL}} u\sqrt{\frac{B}{C}}
\left[R(r,u)\right]^{-1/2}dr \label{Dphiin}
\end{eqnarray}
with $D_{LS}<r_m<D_{OL}$. Even if the function $R(r,u)$ never
vanishes, it becomes minimum at $r=r_m$. Correspondingly, the
integrand has a maximum at this point and is largely dominated by
this peak at $r_m$ if $u$ is very close to $u_m$. So it is
convenient to revisit the analysis of the function $R(r,u)$.

Now we have to keep in mind that $u<u_m$, so that the
parametrization (\ref{ueps}) yields $\epsilon<0$. As pointed
before, the function $R(r,u)$ has no real roots when $u<u_m$, and
in fact, Eq. (\ref{epsdelta2}) with $\epsilon<0$ gives an
imaginary value for $\delta$, so that the inversion point
$r_0=r_m(1+\delta)$ is no longer a real number. Moreover, it is
again convenient to introduce the parametrization (\ref{rzin}),
but this time extended to $r<r_m$ corresponding to $\eta<0$. Thus
now the $\eta$ variable ranges in the interval $1-{r_m\over
r_S}<\eta<1$.

We can now study the function $R(r(\eta),u)$ for small values of $\eta$
as in the previous subsection and find
\begin{eqnarray}
&&R(\eta,u)\simeq R(r_m,u_m)+ \frac{\partial R}{\partial r}
(r_m,u_m) r_m (\eta+\eta^2) \nonumber \\
&&+ \frac{1}{2}\frac{\partial^2 R}{\partial r^2} (r_m,u_m) r_m^2
\eta^2 -\frac{\partial R}{\partial u} (r_m,u_m) u_m |\epsilon|.
\end{eqnarray}
As usual, the first two terms vanish because of Eqs.
(\ref{DRrmum}) and (\ref{Rrmum}). As for the $\epsilon$-term, we
can replace it by the corresponding $\delta^2$-term through Eq.
(\ref{epsdelta}). We just have to keep in mind that now
$\delta^2<0$. The behavior of $R(\eta,u)$ for small $\eta$ is thus
\begin{equation}
R(\eta,u)= \beta_m (-\delta^2+\eta^2)+o(\eta^2).
\end{equation}

Returning to the integral (\ref{Dphiin}), we can change the
integration variable from $r$ to $\eta$ using Eq. (\ref{rzin}) to get
\begin{equation}
\Delta\phi=\int\limits_{\eta_S}^{\eta_O}
u\sqrt{\frac{B(\eta)}{C(\eta)}}\left[R(r,u)\right]^{-1/2}\frac{r_m}{(1-\eta)^2}d\eta
\label{Dphiin1},
\end{equation}
where the integration extrema are already in the form
$\eta_i=1-r_m/D_{Li}$. Note that, since the source is inside the
photon sphere, we have $\eta_S<0$.

As before, we add and subtract a term that contains the main
structure of the integrand, that is the peak at $\eta=0$,
corresponding to $r=r_m$. We have
\begin{eqnarray}
&& \Delta\phi= I_D+I_R \label{Dphiin2}\\
&& I_D= \int\limits_{\eta_S}^{\eta_O}
\frac{u_m}{\sqrt{\beta_m}}\sqrt{\frac{B_m}{C_m}}\frac{r_m}{\sqrt{-\delta^2+\eta^2}}d\eta \label{IDin}\\
&& I_R= \int\limits_{\eta_S}^{\eta_O}
\left[u\sqrt{\frac{B(\eta)}{C(\eta)}}\left[R(\eta,u)\right]^{-1/2}\frac{r_m}{(1-\eta)^2}
\right. \nonumber
\\
&&\left.-\frac{u_m}{\sqrt{\beta_m}}\sqrt{\frac{B_m}{C_m}}\frac{r_m}{\sqrt{-\delta^2
+\eta^2}}\right]d\eta.\label{IRin}
\end{eqnarray}

\begin{figure}
\resizebox{\hsize}{!}{\includegraphics{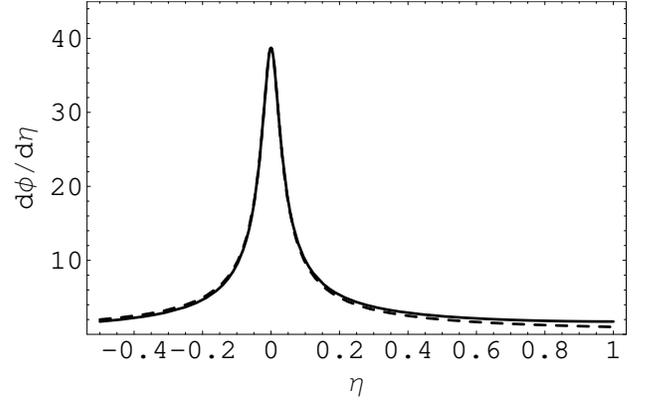}}
\caption{The integrand of $\Delta\phi$ in Eq. (\ref{Dphiin1}) as a
function of $\eta$ for $\delta=0.025 i$ ($\epsilon=-0.001$),
compared to the integrand of $I_D$ in Eq. (\ref{IDin}) in dashed
style. }
 \label{Fig dphidzin}
\end{figure}

The first integral is again elementary and reads
\begin{equation}
I_D=r_m\sqrt{\frac{B_m}{A_m\beta_m}}\log
\frac{\sqrt{-\delta^2+\eta_O^2}+\eta_O}{\sqrt{-\delta^2+\eta_S^2}-|\eta_S|},
\end{equation}
where we have made the sign of $\eta_S$ explicit. The second
integral contains an integrand that is regular everywhere for
$\delta\rightarrow 0$. Fig. \ref{Fig dphidzin} illustrates a
comparison between the integrand of Eq. (\ref{Dphiin1}) and the
integrand of $I_D$ in the Schwarzschild case. We see that even if
there is no divergence for $\eta=0$, the integrand has a very
pronounced peak that dominates the integral. The peak structure is
catched by $I_D$, while the wings are corrected by the
contribution of $I_R$. As $\epsilon \rightarrow 0$, the peak grows
larger and larger, dominating the wings.

Now we make the strong deflection limit approximation, by
requiring that $\epsilon$ and consequently $\delta^2$ is small.
Saving the logarithmically divergent term and the constant terms,
we have
\begin{equation}
\Delta \phi=a \log \frac{4\eta_O\eta_S}{\delta^2}
+b_{SO}+o(\delta^0), \label{Dphiin3}
\end{equation}
where the coefficient $a$ is still given by Eq. (\ref{aSDL}) and
$b_{SO}$ is
\begin{equation}
b_{SO}= \int\limits_{\eta_S}^{\eta_O}g_1(\eta)d\eta \label{b},
\end{equation}
with $g_1(\eta)$ still given by Eq. (\ref{g1}). It is interesting
to note that the argument of the logarithm remains positive, since
both $\delta^2$ and $\eta_S$ are negative.

We can split the integral in Eq. (\ref{b}) in two parts
\begin{eqnarray}
&&b_{SO}=\tilde{b}_{S}+\tilde{b}_{O} \\
&&\tilde{b}_{S}= \int\limits_{\eta_S}^{0}g_1(\eta)d\eta = \int\limits_{0}^{\eta_S}Sign(\eta)\,g_1(\eta)d\eta\label{bSin}\\
&&\tilde{b}_{O}=\int\limits_{0}^{\eta_O}g_1(\eta)d\eta=\int\limits_{0}^{\eta_O}Sign(\eta)\,g_1(\eta)d\eta\label{bOin}
\end{eqnarray}
We observe that $\tilde{b}_{O}$ and $\tilde{b}_{S}$ have the same
formal expression as $b_O$ and $b_S$.

\subsection{Deflection and higher order images formulas}
\label{Sec defl}

As a result of the previous two subsections, we have a unique
expression for the photon deflection, which can be conveniently
stated in terms of $\epsilon$, which represents the impact
parameter shift from the minimum value (see Eq. (\ref{ueps})). By
Eq. (\ref{alpha1SDL}) and (\ref{epsdelta2}) we finally get
\begin{equation}
\Delta \phi=-a\log \frac{\epsilon}{ \eta_O \,\eta_S}+b+\pi,
\label{alphaSDL}
\end{equation}
where we have defined the coefficient
\begin{equation}
b=a\log\frac{2\beta_m}{u_m^2}+b_O+b_S-\pi.
\end{equation}

For quick reference, $\epsilon$ is defined in Eq. (\ref{ueps}),
$u_m$ is given by Eq. (\ref{um}), $\beta_m$ is given by Eq.
(\ref{betam}), $a$ is given by Eq. (\ref{aSDL}),
$\eta_i=1-r_m/D_{Li}$; $b_O$ and $b_S$ are given by the integrals
(\ref{bi}), where $g_1(\eta)$ is found in Eq. (\ref{g1}) and
$R(r,u)$ in Eq. (\ref{Rru}), having changed the integration
variable from $r$ to $\eta$ through the parametrization
(\ref{rzin}).

This expression for the total deflection of the photon is valid
for any position of observer and source. Even if we have not
explicitly considered it, time-reversal symmetry warrants that Eq.
(\ref{alphaSDL}) is valid even in the unrealistic case of an
observer inside the photon sphere (provided that one correctly
relates $\epsilon$ to the sky coordinates of an observer in a
strongly curved region). The only approximation performed is that
the impact parameter $u$ is very close to the critical value
$u_m$.

The general lens equation for spherically symmetric black holes
can be simply stated as
\begin{equation}
\phi_O-\phi_S=\Delta\phi ~~\mathrm{mod}~~ 2\pi.
\end{equation}
Fixing the origin of the azimuthal coordinate in such a way that
$\phi_O=\pi$ and using the expression of the total deflection in
the strong deflection limit (\ref{alphaSDL}), we can easily solve
the lens equation and find the position of the images. In general
we have
\begin{equation}
\epsilon_n=\eta_O\,\eta_S \,e^\frac{b+\phi_S-2n\pi}{a}, \label{epsn}
\end{equation}
where $n$ denotes the number of loops done by the photons before
reaching the observer and $\phi_S\in[-\pi,\pi]$. Of course, the
strong deflection limit becomes exact in the limit $n\rightarrow
\infty$ but is typically a very good approximation already for
$n=1$, as will be shown in the next subsection.

For an observer in the asymptotic region, which is the most
physically interesting case, the angular separation between the
direction of arrival of the photon and the direction of the black
hole is simply $\theta=u/D_{OL}$. So, we have
\begin{eqnarray}
&& \theta=\theta_m(1+\epsilon) \label{thetaeps}\\
&& \theta_m=\frac{u_m}{D_{OL}}. \label{thetam}
\end{eqnarray}
$\theta_m$ is usually called the angular radius of the shadow of
the black hole, since all images of sources outside the photon
sphere reach the observer from angles $\theta>\theta_m$ and the
region within the angular radius $\theta_m$ appears empty.
However, when the source is inside the photon sphere, $D_{LS}<r_m$
and then $\eta_S<0$. We thus have $\epsilon<0$ and the sequence of
images will appear within the shadow of the black hole, with the
lowest order ones closer to the center and the higher order ones
closer and closer to the shadow border.

The study of the Jacobian of the lens equation confirms that the
critical curves are simply given by (\ref{thetaeps}) and
(\ref{epsn}) with $\phi_S=0$ for standard lensing $\phi_S=\pi$ for
retrolensing. They are circles outside the shadow for sources
outside the photon sphere and inside the shadow for sources inside
the photon sphere. The caustics are always pointlike and are
located behind and in front of the source. Altogether, the
caustics cover the whole optical axis as $D_{LS}$ varies from
$+\infty$ to the static limit $r_s$.

It is interesting to take the limit of the total azimuthal shift
$\Delta \phi$ for $D_{OL},D_{LS} \gg r_m$ and calculate the first
order in $r_m/D_{Li}$. Recalling that $\eta_i=1-r_m/D_{Li}$, we
have
\begin{eqnarray}
&&\Delta \phi= -a\log
\frac{u_m^2\epsilon}{2\beta_m}-\frac{ar_m}{D_{OL}}-\frac{ar_m}{D_{LS}}+2\int\limits^{1}_{0}g(\eta)d\eta\nonumber
\\ &&  +g(1)(\eta_O-1)+g(1)(\eta_S-1) +o(\eta_i-1).
\end{eqnarray}
$g(1)$ can be calculated using the asymptotic limit of all metric
functions evaluated in $\eta$ that appear in its expression and
taking the limit for $\eta\rightarrow 1$. It is simply
\begin{equation}
g(1)=\frac{u_m}{r_m}-a.
\end{equation}
Summing up, we get
\begin{equation}
\Delta \phi - \pi =\alpha-\theta_m-\overline{\theta}_m,
\end{equation}
where $\alpha$ is the deflection angle calculated on the
asymptotic trajectories ($D_{LS}=D_{OL}=\infty$), $\theta_m$ is
defined in Eq. (\ref{thetam}) and
\begin{equation}
\overline{\theta}_m=\frac{u_m}{D_{LS}},
\end{equation}
is the angular size of the shadow of the black hole as measured by
a distant source. The lens equation then becomes
\begin{equation}
-\phi_\mathrm{S}=\alpha-\theta_m-\overline{\theta}_m
\end{equation}
The first correction to the lens equation with far source and
observer is thus universal and simply takes into account the
geometry of the lensing problem (compare with the discussion of
the lens equation in Ref. \cite{S2}).

\subsection{Testing the formulas in the Schwarzschild case}

In this subsection, we shall specify all our general formulae for
black hole gravitational lensing in the case of the simplest
possible metric. This will allow us to understand the sense, the
validity and the power of the strong deflection approximation
throughout the range of possible source positions.

The Schwarzschild metric reads (with $G=c=1$)
\begin{eqnarray}
&& A(r)=1-\frac{2M}{r} \\
&& B(r)=\left(1-\frac{2M}{r}\right)^{-1} \\
&& C(r)=r^2.
\end{eqnarray}
The minimum of the function $C/A$ is at $r_m=3M$. Correspondingly
the minimum impact parameter $u_m=\sqrt{C_m/A_m}=3\sqrt{3}M$ as
well-known \cite{Dar,Cha}.

Now let us calculate the coefficients of the deflection formula in
the strong deflection limit. As already noted after Eq.
(\ref{aSDL2}), the coefficient of the logarithmic term is
independent of the source and observer positions. So, Eq.
(\ref{aSDL2}) simply gives the already known value
\begin{equation}
a=1.
\end{equation}

This can be expected, since the logarithmic term is a
characteristic of geodesics winding around the photon sphere which
does not depend on the start and arrival point.

The constant coefficient in the deflection formula is
\begin{eqnarray}
&& b=-\pi+5\log[6]+b_O+b_S \\
&& b_i=-2\log\left[3+\sqrt{3+\frac{18M}{D_{Li}}}\right].
\end{eqnarray}
Putting everything together, we have
\begin{equation}
\Delta \phi -\pi=-\log
\frac{\epsilon}{(1-3M/D_{LS})(1-3M/D_{OL})}+b, \label{alphaSch}
\end{equation}
which reduces to the well-known formula \cite{Dar}
\begin{equation}
\alpha=-\log \epsilon+\log[216(2-\sqrt{3})^2]-\pi
\label{alphaSchinf}
\end{equation}
in the limit $D_{OL},D_{LS}\rightarrow \infty$. Eq.
(\ref{alphaSch}) is the generalization of Darwin's formula
(\ref{alphaSchinf}) to sources and observers at finite distance
from the black hole, the only approximation remaining $|\epsilon|
\ll 1$ (see Appendix \ref{AppB} for a discussion of formulae
expressed in terms of alternative perturbative parameters). In
order to test our formula for the deflection of a photon, we can
use it to calculate the radius of the critical curves.

\begin{figure}
\resizebox{\hsize}{!}{\includegraphics{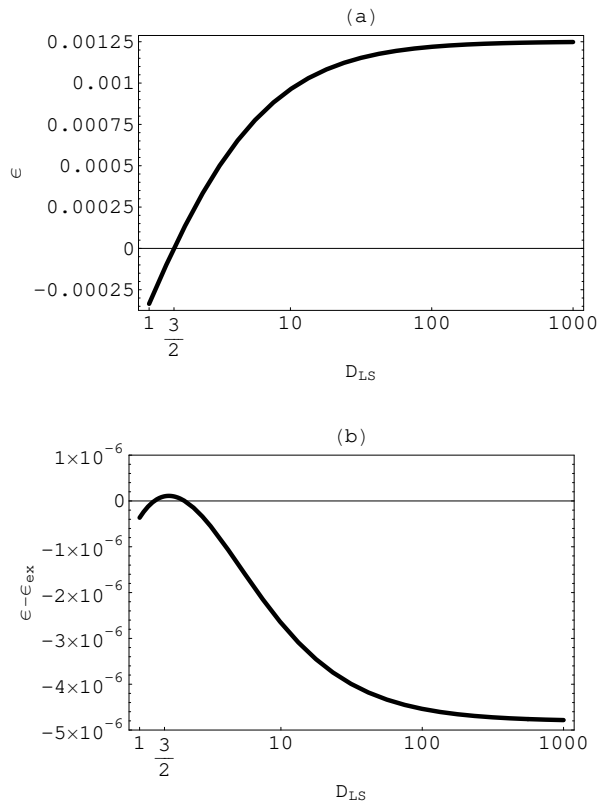}} \caption{(a)
Angular radius of the first relativistic Einstein ring relative to
the black hole shadow for a source behind the black hole as a
function of the source distance measured in Schwarzschild radii
($\epsilon_3=(\theta_3-\theta_m)/\theta_m$, $2M=1$). (b)
Difference between the value of $\epsilon_3$ calculated by the
strong deflection limit formula and the one calculated by the
exact formula.}
 \label{Fig eps}
\end{figure}

The angular radius of the critical rings is given by Eq.
(\ref{epsn}) with $\phi_S=0$ or $\pi$. Explicitly, for the
Schwarzschild metric we have
\begin{eqnarray}
&&\theta_k=\frac{3\sqrt{3}M}{D_{OL}}(1+\epsilon_k) \label{CritSch}\\
&& \epsilon_k=
\frac{216(2-\sqrt{3})(2D_{LS}-3)}{(\sqrt{3D_{LS}}+\sqrt{3+D_{LS}})^2}e^{-k\pi},
\label{epsk}
\end{eqnarray}
where $k$ is an even number in the retro-lensing case
($\phi_S=\pi$), and an odd number in the standard lensing case
($\phi_S=0$). The first critical curve for $k=1$ is created by
photons experiencing weak deflection and is beyond the range of
validity of Eq. (\ref{CritSch}). The critical curve with $k=2$ is
the first retro-lensing ring, while for $k=3$ we obtain the first
higher order Einstein ring of standard lensing.

The displacement of the first relativistic Einstein ring from the
black hole shadow is shown in Fig. \ref{Fig eps}a, where it can be
appreciated how $\epsilon_3$ stays small throughout the range of
source distances, validating the strong deflection limit
approximation. For $D_{LS}\gg M$, the ring radius tends to its
asymptotic value. Our analytic formula for the critical curve
nicely joins the region within the photon sphere $D_{LS}<3M$ to
that outside the photon sphere $D_{LS}>3M$. The divergent term in
the deflection formula at $D_{LS}=3M$ forces $\epsilon=0$, in
order to keep $\Delta \phi$ finite. Since this is true for any
order $k$, we can conclude that all higher order images of a
source right at the photon sphere collapse into one degenerate
image appearing right at the shadow border. As $D_{LS}<3M$,
$\epsilon$ becomes negative as expected. This means that the
sequence of Einstein rings for a source inside the photon sphere
is reversed: the brightest rings will appear closer to the center
of the shadow and the fainter will be farther, approaching the
shadow border as $k$ grows more and more.

The Schwarzschild metric is simple enough to allow an exact
integration of the azimuthal motion in terms of elliptic
integrals. It is thus very important and instructive to compare
the critical curve radius calculated by the formula
(\ref{alphaSDL}) obtained in the strong deflection limit to the
exact Einstein ring position, calculated using the exact
deflection angle. The difference between the two positions is
plotted in Fig. \ref{Fig eps}b, where it can be appreciated that
it stays of the order $\epsilon^2$ throughout the range of
$D_{LS}$, testifying the accuracy of the strong deflection limit
as a powerful approximation to describe higher order images.

The updated analysis of extreme gravitational lensing by
Schwarzschild black holes presented here can be repeated for any
kind of spherically symmetric black holes, using the formulae
derived in this section. It is interesting to consider that some
metrics could give the same lensing observables for sources at
infinity, whereas they could be distinguished in gravitational
lensing of sources at small distances from the black hole.
Therefore, the introduction of $D_{LS}$ as a new parameter
enriches the arena for the comparison of different metrics.

\section{Kerr black hole} \label{Sec Kerr}

In this section, we shall describe gravitational lensing of
sources placed at arbitrary distances from a spinning black hole.
With respect to the analysis of Ref. \cite{KerGen}, the finiteness
of $D_{OL}$ and $D_{LS}$ only intervenes in the calculation of the
radial integrals. As a consequence, the first three sections of
Ref. \cite{KerGen}, concerning the description of the unstable
circular orbit and the shadow of the Kerr black hole, remain
unaffected. Their content is briefly reviewed in the following
subsection. The modifications in the radial integrals are
reflected in the lens equation and its Jacobian. Consequently,
also the critical curves and the caustics contain a dependence on
the source position. They are described in Section \ref{Sec Kerr
Caustics}. Finally, lensing of sources near caustics is updated in
Section \ref{Sec Kerr lensing}. Throughout this section, we shall
preserve the spirit of Ref. \cite{KerGen} expanding all quantities
to the second order in the black hole spin $a$. The perturbative
expansion is of key-importance to keep all calculations fully
analytic up to the final results. At the same time, the second
order approximation proves to be very reliable up to values of the
black hole spin $a\simeq 0.1$, as noted in the comparison with
numerical results \cite{KerObs}.

\subsection{Derivation of the lens equation} \label{Sec GenOut}

The Kerr metric in Boyer-Lindquist coordinates \cite{BoyLin} is
\begin{eqnarray}
& ds^2=&\frac{\Delta-a^2 \sin^2 \vartheta}{\rho^2}d
t^2-\frac{\rho^2}{\Delta} dr^2-\rho^2 d\vartheta^2 \nonumber \\
&& - \frac{ \left(r^2+a^2 \right)^2 - a^2\Delta \sin^2 \vartheta
}{\rho^2} \sin^2 \vartheta d\phi^2 \nonumber \\&&+\frac{2ar
\sin^2\vartheta}{\rho^2} dt d\phi \\%
& \Delta=&r^2-r+a^2, \\%
& \rho^2=& r^2+a^2 \cos^2\vartheta.
\end{eqnarray}
Distances are measured in units of the Schwarzschild radius
($2MG/c^2=1$) and $a$ is the specific angular momentum of the
black hole, running from $0$ (Schwarzschild black hole) to $1/2$
(extremal Kerr black hole) in our units.

The Kerr geodesics are described in integral form by the equations
\begin{eqnarray}
&& \pm \int \frac{dr}{\sqrt{R}}=\pm \int \frac{d
\vartheta}{\sqrt{\Theta}} \label{Geod1}\\
& \phi_O-\phi_S =& a \int\frac{r^{2}+a^{2}-a J}{\Delta \sqrt{R}}
dr-a \int \frac{dr}{\sqrt{R}} \nonumber  \\
&& + J \int \frac{\csc^2\vartheta}{\sqrt{\Theta}} d \vartheta,
\label{Geod2}
\end{eqnarray}
with
\begin{eqnarray}
&\Theta=&Q+a^2 \cos^2\vartheta-J^2 \cot^2\vartheta \\
&R=&r^4+(a^2-J^2-Q)r^2+(Q+(J-a)^2) r \nonumber \\ &&-a^2 Q.
\label{R}
\end{eqnarray}
$J$ is the component of the angular momentum of the photon along
the spin axis and $Q$ is the Carter integral \cite{Car}, related
to the total angular momentum of the photon (we set the specific
energy $E$ to 1 by a suitable choice of the affine parameter).

An observer in the position $(D_{OL},\vartheta_O,\phi_O)$, defines
angular coordinates $(\theta_1,\theta_2)$ in the sky, such that
the black hole is in $(0,0)$ with its spin projected along
$\theta_2$. A photon travelling on a geodesic characterized by
constants of motion $J$ and $Q$ hits a distant observer from the
direction
\begin{eqnarray}
&& \theta_{1}=- \frac{J}{D_{OL} \sqrt{1-\mu_O^2}}, \label{Th1J} \\
&& \theta_{2}=\pm D_{OL}^{-1}\sqrt{Q + \mu_O^2\left( a^2-
\frac{J^2}{1-\mu_O^2} \right) },\label{Th2Q}
\end{eqnarray}
where $\mu_O\equiv \cos \vartheta_O$ as usual \cite{Cha}.

\subsubsection{Shadow of the Kerr black hole}

Among all photon trajectories ending at the observer, there is a
family of trajectories that approach an unstable circular orbit
around the black hole when traced back asymptotically in the past.
This family can be parameterized by the parameter $\xi$ ranging
from $-1$ to 1. The constants of motion identifying the geodesics
of this family are then given by
\begin{eqnarray}
&J_m(\xi)=&\frac{3\sqrt{3}}{2}\xi\sqrt{1-\mu_O^2}-a(1-\mu_O^2)(1+\xi^2)
\nonumber
\\ && -a^2 \frac{\xi \sqrt{1-\mu_O^2}}{3\sqrt{3}}[5-2 \xi^2-2
\mu_O^2(1-\xi^2)], \label{Jma2}
\end{eqnarray}
\begin{eqnarray}
&Q_m(\xi)=&\frac{27}{4}\left[ 1-(1-\mu_O^2)\xi^2 \right] \nonumber \\
&&
-3\sqrt{3}a\xi \sqrt{1-\mu_O^2}[1+\mu_O^2-(1-\mu_O^2)\xi^2] \nonumber \\
&& -a^2[(1+\mu_O^2)^2-4(1-\mu_O^2)\xi^2 \nonumber \\
&& +3(1-\mu_O^2)^2\xi^4]. \label{Qma2}
\end{eqnarray}

The radius of the unstable circular orbit asymptotically
approached in the past is
\begin{eqnarray}
&r_{m}=& \frac{3}{2}-\frac{2}{\sqrt{3}}a\xi
\sqrt{1-\mu_O^2}-\frac{4}{9}a^2(1+\mu_O^2) \nonumber \\ &&
-\frac{4}{27\sqrt{3}} a^3\xi(5+6\mu_O^2)\sqrt{1-\mu_O^2}+O(a^4).
\label{xmxi}
\end{eqnarray}
Correspondingly, a distant observer sees such photons from the
directions
\begin{eqnarray}
&D_{OL} \theta_{1,m}&=-\frac{3\sqrt{3}}{2}\xi+a
\sqrt{1-\mu_O^2}(1+\xi^2) \nonumber \\ &&
+a^2\frac{\xi}{3\sqrt{3}}[5-2\mu_O^2 -2 \xi^2(1-\mu_O^2)],
\label{Th1m}
\end{eqnarray}
\begin{eqnarray}
&D_{OL} \theta_{2,m}& = \pm\frac{3\sqrt{3}}{2}\sqrt{1-\xi^2} \mp
a\xi \sqrt{1-\xi^2} \sqrt{1-\mu_O^2} \nonumber \\ &&  \mp a^2
\frac{\sqrt{1-\xi^2}}{3\sqrt{3}}[1+2\mu_O^2-2\xi^2(1-\mu_O^2)].
\label{Th2m}
\end{eqnarray}

Eqs. (\ref{Th1m}) and (\ref{Th2m}) define a curve in the observer
sky as $\xi$ varies between $-1$ and $1$. This curve represents
the border of the shadow of the black hole, in the sense that all
photons emitted by a source outside the unstable circular orbits
reach the observer from directions outside this border. In the
Schwarzschild limit, the shadow border is simply a circle of
radius $3\sqrt{3}/2D_{OL}$, whereas for generic values of the
spin, the shadow satisfies the ellipse equation
\begin{equation}
\frac{(\theta_{1,m}-\theta_0)^2}{A_1^2}+
\frac{\theta_{2,m}^2}{A_2^2}=1 +o(a^2)\label{shadow}
\end{equation}
with the origin shifted rightward by
\begin{equation}
\theta_0=\frac{2a\sqrt{1-\mu_o^2}}{D_{OL}}, \label{ShadowShift}
\end{equation}
and semiaxes given by
\begin{eqnarray}
&& A_1=\frac{3\sqrt{3}}{2D_{OL}} \left(1-\frac{2}{9}a^2 \right) \\
&& A_2=\frac{3\sqrt{3}}{2D_{OL}}
\left(\frac{3\sqrt{3}}{2}-\frac{2\mu_o^2}{9}a^2 \right),
\end{eqnarray}
with ellipticity
\begin{equation}
e\equiv 1-\frac{A_1}{A_2}=\frac{2}{9}a^2(1-\mu_o^2)
\label{ellshadow}
\end{equation}

For more details on the derivation of the parameterization
(\ref{xmxi}) and of the shadow border, see the deep and detailed
discussion in Ref. \cite{KerGen}.

\subsubsection{Strongly deflected photons}

We now introduce the following parametrization of the observer sky
by
\begin{equation}
\left\{ \begin{array}{l}
 \theta_1(\epsilon,\xi)=\theta_{1,m}(\xi)(1+\epsilon) \\
\theta_2(\epsilon,\xi)=\theta_{2,m}(\xi)(1+\epsilon)
\end{array}
\right. . \label{ThetaParam}
\end{equation}
One half of the sky is covered as $\xi$ varies from $-1$ to $1$
and $\epsilon$ varies from $-1$ to $+\infty$. The double sign in
$\theta_{2,m}$ selects which half of the sky we are covering. We
are interested into strongly deflected photons, which correspond
to very small values of $\epsilon$. One may regard $\epsilon$ as
the relative displacement of the photon direction from the shadow
border. As $\epsilon \rightarrow 0$, the photon spends more and
more time close to the unstable circular orbit, and performs more
and more loops around the black hole before emerging.

Using Eqs. (\ref{Th1J}) and (\ref{Th2Q}) we can calculate the
values of $J$ and $Q$ identifying the geodesics of such photons.
For each value of $J$ and $Q$, we can calculate the inversion
point $r_0$ in the radial motion examining the roots of the
function $R(r,J,Q)$, defined by Eq. (\ref{R}). It is immediate to
see that $R(r,J,Q)$ satisfies the properties
\begin{eqnarray}
&& \frac{\partial R}{\partial r}(r_m,J_m,Q_m)=0 \label{RrmJmQm}\\
&& R(r_m,J_m,Q_m)=0  \\
&& R(r_0,J,Q)=0. \label{Rr0JQ}
\end{eqnarray}

Analogously to the spherically symmetric case, we can define the
parameter $\delta$ by
\begin{equation}
r_0=r_m(1+\delta).
\end{equation}
$\delta$ is thus the relative displacement of the inversion point
from the unstable circular orbit radius. Inserting the expression
of $J$ and $Q$ as functions of $\xi$ and $\epsilon$ and solving
Eq. (\ref{Rr0JQ}) for $\delta$ to the lowest order in $\epsilon$
we find that the two parameters are related by
\begin{equation}
\delta = \sqrt{\frac{2\epsilon}{3}}\left[ 1
-\frac{2}{3\sqrt{3}}a\hat\xi +\frac{2}{27}a^2 (10-\mu_o^2-14
\hat\xi^2)\right], \label{deltatoeps}
\end{equation}
where we have introduced the compact notation
\begin{equation}
\hat\xi=\xi \sqrt{1-\mu_o^2}. \label{xindef}
\end{equation}

So, for any photons hitting the observer from a direction close to
the shadow border, we can immediately find the corresponding value
of the inversion point in its motion around the black hole. If the
photon comes from the interior of the shadow ($\epsilon<0$), then
the inversion point becomes complex, signaling the fact that the
photon follows a geodesic without inversion points in the radial
motion.

\subsubsection{From the geodesics equation to the lens equation}

Now we can come to the resolution of the radial integrals
appearing in the geodesics equations (\ref{Geod1}) and
(\ref{Geod2}). We indicate them by
\begin{eqnarray}
&& I_1= \pm\int \frac{dr}{\sqrt{R}} \label{I1}\\
&& I_2=\pm\int \frac{r^{2}+a^{2}-a J}{\Delta \sqrt{R}} dr
\label{I2}.
\end{eqnarray}

These integrals must be performed along the photon path from the
source to the observer. The double sign indicates that one has to
sum the contributions with different signs of $\dot r$ coherently.
The task is simplified by the observation that the function
$R(r,J,Q)$ determining the radial motion in the Kerr metric
satisfies the same properties as the function $R(r,u)$ determining
the radial motion in spherically symmetric metrics (compare Eqs.
(\ref{RrmJmQm})-(\ref{Rr0JQ}) to Eqs.
(\ref{DRrmum})-(\ref{Rr0u})). So, we can repeat exactly the same
steps of the analysis of the function $R(r,u)$, with the trivial
extension to the presence of two constants of motion. The
integrals are then easily solved to
\begin{eqnarray}
& I_1=&- a_1 \log \delta+ b_1+c_1(D_{LS})+c_1(D_{OL}) \label{I1solved}\\
& I_2=&- a_2 \log \delta+ b_2+c_2(D_{LS})+c_2(D_{OL}),
\label{I2solved}
\end{eqnarray}
where the coefficients $a_1$, $b_1$, $a_2$, and $b_2$ remain the
same as those given in the appendix of Ref. \cite{KerGen}, whereas
the explicit expressions of the new functions $c_1$ and $c_2$ are
shown in the Appendix \ref{AppA} at the end of this work (Eqs.
(\ref{c1}) and (\ref{c2})). We have stored the whole dependence on
the source and observer distance in these two coefficients. They
are defined in such a way that they vanish when their argument
goes to infinity, so that the expressions for distant sources and
observers are recovered. Although we have provided the expressions
of the integrals for arbitrary observer distances, we shall
consider $D_{OL}\gg 1$ from now on for simplicity. If one is
interested to study gravitational lensing with observers close to
the black hole, it is easy to recover the relevant formulae, since
the dependence on $D_{LS}$ and $D_{OL}$ is interchangeable, thanks
to the symmetry $D_{LS}\leftrightarrow D_{OL}$ in Eqs.
(\ref{I1solved}) and (\ref{I2solved}).

In the angular integrals
\begin{eqnarray}
&& J_1=\pm \int \frac{1}{\sqrt{\Theta}} d \vartheta \label{J1} \\
&& J_2=\pm \int \frac{csc^2\vartheta}{\sqrt{\Theta}} d \vartheta
\label{J2}
\end{eqnarray}
there is no dependence on the source and observer distance.
Therefore, we can safely exploit the expressions in the appendix
of Ref. \cite{KerGen} without any more concern.

Finally, it is convenient to replace the parameter $\delta$
(which, as we recall, represents the relative displacement of the
inversion point from the radius of the unstable circular orbit) by
a new variable
\begin{equation}
\psi=-2\log\delta+\log\left[\frac{144(7-4\sqrt{3})(2D_{LS}-3)}{(\sqrt{3D_{LS}}+\sqrt{3+D_{LS}})^2}\right].
\label{psidelta}
\end{equation}

From the physical point of view, $\psi$ simply represents the
equivalent azimuthal shift of a photon deflected by a
Schwarzschild black hole.

Once all integrals are solved, we can rearrange the integrated
geodesics equations in the typical form of a lens equation, moving
the source angular coordinates on the left hand side and leaving
everything else on the right hand side
\begin{equation} \left\{
\begin{array}{l}
\mu_S=\mu_S(\psi,\xi;\mu_O,D_{OL},D_{LS})  \\
\phi_S=\phi_S(\psi,\xi;\mu_O,D_{OL},D_{LS})
\end{array} \right. . \label{LensApp}
\end{equation}
The lens mapping relates the source coordinates $(\mu_S,\phi_S)$
to the set of intermediate variables $(\psi,\xi)$, which
characterize the photon geodesic by the amount of deflection and
its orientation in space respectively. The sky coordinates
$(\theta_1,\theta_2)$ are related to $(\psi,\xi)$ by Eqs.
(\ref{ThetaParam}), (\ref{deltatoeps}) and (\ref{psidelta}). The
observer position and the source distance play the role of
parameters of the lens mapping.

In the following subsections we shall put in evidence the main
features of the lens mapping through the analysis of its critical
points and the discussion of gravitational lensing near the
critical points.
%

\subsection{Critical curves and caustics} \label{Sec Kerr Caustics}

In order to find the critical points of the lens equation, one has
to calculate the Jacobian determinant
\begin{equation}
\det J= \frac{\partial \mu_s}{\partial \xi} \frac{\partial
\phi_s}{\partial \psi}-\frac{\partial \mu_s}{\partial
\psi}\frac{\partial \phi_s}{\partial \xi}. \label{DetJ}
\end{equation}
The critical points of the lens mapping are the solutions of the
equation $\det J(\psi,\xi)=0$. The critical curves are the
corresponding points in the observer sky $(\theta_1,\theta_2)$
through Eq. (\ref{ThetaParam}) and the caustics are the images of
these points in the $(\mu_s,\phi_s)$ space through the lens
mapping. The full procedure is straightforward and is detailed in
Ref. \cite{KerGen}. Here we just state the updated results with
the encompassment of the finiteness of $D_{LS}$.

\subsubsection{Critical points in the $(\psi,\xi)$ space}

As mentioned before, $\psi$ represents the equivalent azimuthal
shift of a photon moving in the Schwarzschild metric obtained
turning the black hole spin off. It is thus not surprising that
the zero-order critical points are simply given by
$\psi_{k}^{(0)}=k\pi$, with $k$ being an integer number. $k$ is
the number of half-loops performed by a photon in the
Schwarzschild metric. We will often refer to this integer number
as the critical curve order or caustic order. $k=1$ gives the
azimuthal shift of a photon emitted by a source behind the black
hole and weakly deflected by the black hole. Such photons are not
the subject of our analysis. The first interesting case is $k=2$
corresponding to photons emitted by a source in front of the black
hole and backscattered by the black hole (retro-lensing). Photons
with $k=3$ are again emitted by a source behind the black hole,
but now the photons perform a complete loop around the black hole
before reaching the observer. Summing up, odd critical orders are
involved in gravitational lensing of a source behind the black
hole, whereas even critical orders are involved in retro-lensing.
For each $k$ we have a different critical curve, which physically
corresponds to the degenerate image of a source placed on the
corresponding caustic. Starting by the zero-order solution, the
first and second order correction can be found solving the
Jacobian equation order by order. For each $k$, we have
\begin{equation}
\psi_{k}(\xi)= k\pi+ a \psi_{k}^{(1)}(\xi)+ a^2
\psi_{k}^{(2)}(\xi), \label{psicr}
\end{equation}
with $\psi_{k}^{(1)}$ and $\psi_{k}^{(2)}$ given by
\begin{eqnarray}
&\psi_{k}^{(1)}&= \frac{2\hat\xi\sqrt{D_{LS}}}{(2D_{LS}-3)\sqrt{3+D_{LS}}} \label{psicr1}\\
&\psi_{k}^{(2)}&= -\frac{1}{18} \left[9c_k
(3-2\mu_O^2-3\hat\xi^2)+32(1-\hat\xi^2)\right] \nonumber \\
&& -\left[3\sqrt{3}
(2D_{LS}-3)^2\sqrt{D_{LS}}(3+D_{LS})^{3/2}\right]^{-1}
\nonumber \\
&& \cdot \left\{ (2D_{LS}-3)(3+D_{LS})\left[2
\left(\sqrt{D_{LS}}-\sqrt{3+D_{LS}}\right)^2 \right. \right. \nonumber \\
&&\cdot (2D_{LS}-3)(3-2\mu_O^2-3\hat\xi^2) \nonumber \\
&& \left. +26D_{LS}-27-(4D_{LS}-18)\mu_O^2-(18D_{LS}-27)\hat\xi^2
\right] \nonumber \\
&& \left. -108D_{LS}\hat\xi^2 \right\}.\label{psicr2}
\end{eqnarray}
In the limit $D_{LS}\rightarrow \infty$ the first order correction
vanishes while the second order correction reduces to the first
row. These correctly reproduce the results of Ref. \cite{KerGen}.
As $\xi$ varies between $-1$ and 1, we recover all critical points
in the $(\psi,\xi)$ space.

\subsubsection{Critical curves}

Starting from the solution of the Jacobian determinant equation in
the $(\psi,\xi)$ space, we can construct both the critical curves
and the caustics. As for the critical curves, it is sufficient to
use Eq. (\ref{ThetaParam}) with $\epsilon$ expressed by Eqs.
(\ref{deltatoeps}) and (\ref{psidelta}) in terms of $\psi$ and put
$\psi=\psi_{k}$ as given by Eq. (\ref{psicr}). In this way one
gets the critical curves in the form
$(\theta_{1,k}(\xi),\theta_{2,k}(\xi))$. These expressions are
lengthy and are not very transparent. However, it is
straightforward to prove that they satisfy the ellipse equation at
the second order in $a$
\begin{equation}
\frac{(\theta_{1,k}-\theta_{0,k})^2}{A_{1,k}^2}+
\frac{\theta_{2,k}^2}{A_{1,k}^2}=1 +o(a^2). \label{CritEllipse}
\end{equation}

\begin{figure}
\resizebox{8cm}{!}{\includegraphics{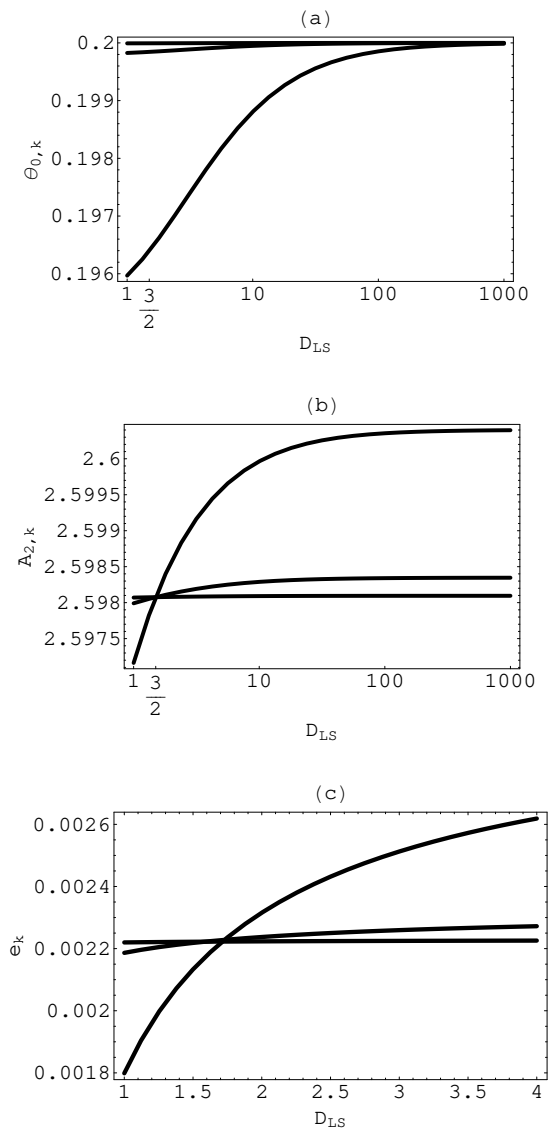}} \caption{(a)
Apparent shift of the center of the critical curves with respect
to the black hole position as a function of the source distance;
from the bottom up, the curves are for $k=2,3,4$, respectively.
(b) Major semiaxis of the critical curves with $k=2,3,4$ (from top
to bottom) as a function of the source distance. (c) Ellipticity
of the critical curves as a function of the source distance; from
the top down the curves are done for $k=2,3,4$, respectively. All
plots are done for $a=0.1$ and equatorial observer $\mu_O=0$.}
 \label{Fig Crit}
\end{figure}

The center of the critical curve is shifted by the quantity
\begin{equation}
D_{OL}\theta_{0,k}=2a\sqrt{1-\mu_o^2}
\left(1-\frac{3\sqrt{3D_{LS}}\epsilon_k}{(2D_{LS}-3)\sqrt{3+D_{LS}}}
 \right), \label{CritShift}
\end{equation}
which reduces to the shadow shift in the limit
$D_{LS}\rightarrow\infty$. For generic values of the source
distance, the center of the critical curve does not coincide with
the center of the shadow. However, the displacement is very small,
since it is proportional to $\epsilon_k$ (given in Eq.
(\ref{epsk}), which is very small in the strong deflection limit.
We can also note that the degeneracy between the black hole spin
$a$ and the observer position $\sqrt{1-\mu_o^2}$ that was pointed
out in Ref. \cite{KerGen} holds even when the source is at finite
distance. Fig. \ref{Fig Crit}a shows the dependence of the shift
with $D_{LS}$. For any value of the critical order $k$, the shift
tends to the shadow shift for large values of $D_{LS}$. The first
retrolensing critical curve ($k=2$) is the most displaced one.

The major semiaxis of the critical curve is oriented along the
projection of the spin on the observer sky. Its explicit
expression is
\begin{eqnarray}
&A_{2,k}&=\frac{3\sqrt{3}}{2D_{OL}}+\frac{a^2}{D_{OL}}
\left\{-\frac{\mu_O^2}{\sqrt{3}}+\epsilon_k\frac{4\mu_O^2-9c_k(2\mu_O^2-3)}{4\sqrt{3}}
\right.
\nonumber \\
&&+\epsilon_k\left[2\sqrt{D_{LS}}\sqrt{3+D_{LS}}(2D_{LS}-3)\right]^{-1}
\nonumber \\
&&
\left[2(2D_{LS}-3)(\sqrt{D_{LS}}-\sqrt{3+D_{LS}})^2(3-2\mu_O^2)\right.
\nonumber \\ &&\left. \left.+9
(2D_{LS}-3)+2\mu_O^2(2D_{LS}+9)\right] \right\},
\end{eqnarray}
and is plotted in Fig. \ref{Fig Crit}b as a function of the source
distance. We can appreciate that all critical curves are external
to the shadow as $D_{LS}>3/2$ and internal to it when
$D_{LS}<3/2$, as in the spherically symmetric case. The larger the
order of the critical curve, the closer the curve is to the
shadow, as $\epsilon_k\rightarrow 0$ for $k\rightarrow \infty$.

Rather than giving the expression of the minor semiaxis, it is
instructive to examine the ellipticity of the critical curve,
defined by
\begin{equation}
e_k=1-\frac{A_{1,k}}{A_{2,k}}.
\end{equation}
Of course, $A_{1,k}$ can be easily derived by this expression if
one is interested in it. To the second order in $a$ and to the
first order in $\epsilon_k$, the ellipticity is
\begin{eqnarray}
&e_k&=a^2(1-\mu_O^2)\left\{\frac{2}{9}+\frac{(27c_k-8)\epsilon_k}{18}
\right.
\nonumber \\
&&-\frac{4\epsilon_k}{\sqrt{3}}+\epsilon_k\left[9\sqrt{D_{LS}}(3+D_{LS})^{3/2}(2D_{LS}-3)^2
\right]^{-1} \nonumber \\
&& \cdot
\left[\sqrt{3}(729-225D_{LS}-348D_{LS}^2+92D_{LS}^3+48D_{LS}^4)\right. \nonumber \\
&& \left.\left.-72D_{LS}^{3/2}\sqrt{3+D_{LS}}\right] \right\}.
\label{ellcr}
\end{eqnarray}
This quantity reduces to the first row when $D_{LS}\rightarrow
\infty$, which is the same as that given in Ref. \cite{KerGen}
safe for the fact that we stop here at the first order in
$\epsilon_k$ in order to be consistent with the strong deflection
limit approximation. For large values of $k$,
$\epsilon_k\rightarrow 0$ and the ellipticity of the critical
curves tend to the ellipticity of the shadow (\ref{ellshadow}).
Fig. \ref{Fig Crit}c shows the ellipticity of the first three
relativistic critical curves as functions of the source distance.
We note that all curves are more elliptical than the shadow when
$D_{LS}$ is large. But for sources just slightly farther than the
unstable circular photon orbit, the ellipticity of the critical
curves equals the ellipticity of the shadow. This happens at a
value of $D_{LS}$ slightly greater than $3/2$ and different for
all critical curves. Below this distance, the ellipticity of the
critical curves becomes smaller than that of the shadow. Finally,
we can note that the ellipticity remains a function of
$a\sqrt{1-\mu_O^2}$, thus preserving the degeneracy between the
black hole spin and its orientation relative to the line of sight.

\subsubsection{Caustics}

The caustics are obtained evaluating the lens mapping in the
critical points $\psi_k$ given by Eq. (\ref{psicr}). To the zero
order, the Schwarzschild caustics are recovered. They are
pointlike and placed behind the black hole for $k$ odd and in
front of the black hole for $k$ even. As the black hole spin is
turned on, the caustics drift from the optical axis and acquire a
finite extension. Their explicit expression is
\begin{eqnarray}
&& \mu_S= (-1)^k\mu_O \pm R_k\sqrt{1-\mu_O^2} (1-\xi^2)^{3/2},
\label{Caumu2} \\
&& \phi_S=(1-k)\pi-\Delta \phi_k-\frac{R_k}{ \sqrt{1-{\mu_o}^2}}
\xi^3, \label{Caugamma2}
\end{eqnarray}
with the drift given by
\begin{eqnarray}
&\Delta\phi_k&=a\left\{\frac{4k\pi}{3\sqrt{3}}+2\log\left[3(2-\sqrt{3})^2\right]
\right. \nonumber \\
&&\left. +\log\left[
\frac{(2\sqrt{D_{LS}}+\sqrt{3+D_{LS}})^2}{9(D_{LS}-1)}\right]
\right\}, \label{Deltaphik}
\end{eqnarray}
and the semi-amplitude given by
\begin{eqnarray}
&R_k&=a^2(1-\mu_O^2)\left[c_k \right. \nonumber \\
&& \left. +\frac{2\left(
9+4D_{LS}-4\sqrt{D_{LS}}\sqrt{3+D_{LS}}\right)}{3\sqrt{3}\sqrt{D_{LS}}\sqrt{3+D_{LS}}}\right]
\label{Rk}
\end{eqnarray}

\begin{figure}
\resizebox{\hsize}{!}{\includegraphics{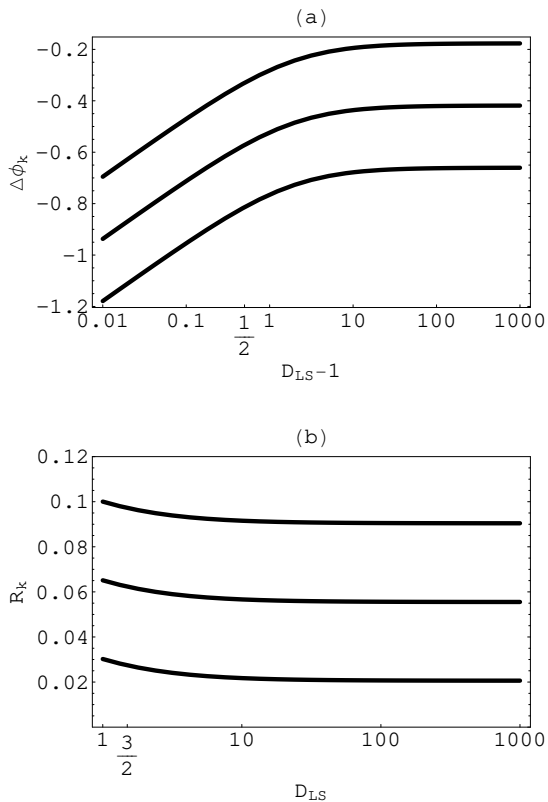}} \caption{(a)
Drift of the center of the caustic from the optical axis as a
function of the source distance; from the bottom up, the curves
are for $k=2,3,4$, respectively. (b) Semiamplitudes of the
caustics with $k=2,3,4$ (from bottom to top) as a function of the
source distance. Both plots are done for $a=0.1$ and equatorial
observer $\mu_O=0$.}
 \label{Fig Cau}
\end{figure}

The caustic is a four-cusped astroid with the same angular
extension along both axes, as can be explicitly seen transforming
the above expressions to coordinates centered on the caustic (see
Ref. \cite{KerGen}). The outstanding feature that emerges from
these expressions is that the drift of the caustic diverges
logarithmically as the source approaches the horizon. As can be
seen in Fig. \ref{Fig Cau}a, the drift is always negative
(clockwise as seen from the northern pole of the black hole) and
grows linearly with the caustic order $k$. The drift tends to the
asymptotic value (given by the first row of Eq. (\ref{Deltaphik}))
for large values of $D_{LS}$, while it grows monothonically as
$D_{LS}$ is decreased.

The amplitude of the caustic does not present any divergences. As
shown in Fig. \ref{Fig Cau}b, the amplitude increases linearly
with the caustic order and tends to the asymptotic value (given by
the first row in Eq. (\ref{Rk}) for large $D_{LS}$. As $D_{LS}$ is
decreased up to the horizon it grows by a fixed amount, given by
\begin{equation}
R_k(1)-R_k(\infty)=\frac{5}{3\sqrt{3}}a^2(1-\mu_O^2).
\end{equation}

Taking $D_{LS}$ as a parameter ranging from $1$ to $+\infty$, we
can trace the whole caustic hypersurface using Eqs. (\ref{Caumu2})
and (\ref{Caugamma2}). The result is shown in Fig. \ref{Fig
Cau3D}, where the caustic appears as a tube with the transverse
section having the shape of a four cusped astroid. At large
distances the caustic tube keeps its angular extension fixed and
thus covers a larger and larger area. Close to the horizon, the
caustic tube winds around the black hole indefinitely. A similar
picture has already been done by Rauch \& Blandford \cite{RauBla}
by numerical techniques for the caustic of order $k=1$. Our plot
is entirely based on our analytical formulae (\ref{Caumu2}) and
(\ref{Caugamma2}), which are valid for arbitrary order except
$k=1$. Our study is thus complementary to that of Ref.
\cite{RauBla}. The origin of the logarithmic divergence in the
caustic angular position can be traced back to the divergence of
the integral $I_2$ in Eq. (\ref{Geod2}) (see also Eq. (\ref{c2})
in Appendix \ref{AppA}). Indeed, the integrand contains a factor
$\Delta^{-1}$, which diverges linearly as the integration variable
$r$ approaches the horizon. Since $D_{LS}$ is the lower bound of
that integral, $I_2$ must diverge logarithmically as $D_{LS}$
tends to the horizon. The divergence of $I_2$ has no effect in
Schwarzschild, since it appears multiplied by the black hole spin,
but as soon as $a\neq 0$, the logarithmic divergence is
transferred to the azimuthal shift $\phi_O-\phi_S$, so that
photons emitted by a source very close to the horizon must perform
several loops before exiting. The latter argument is completely
independent of our perturbative expansion in the black hole spin,
proving that the logarithmic divergence is not an artifact of our
perturbative framework. Moreover, the divergence of $I_2$ is not
even a product of the strong deflection limit, since it is still
there whatever the values of the constants of motion $J$ and $Q$.
Therefore, it seems plausible to us that even the primary caustic
tube (with $k=1$), which is not included in our treatment, should
show a similar behavior, winding an infinite number of times
before entering the horizon. This seems not to be observed in the
numerical analysis by Rauch \& Blandford \cite{RauBla}, where the
primary caustic tube always appears to perform a finite number of
loops before plunging into the horizon, except for the extremal
case $a=0.5$. It is anyhow difficult to believe that the
divergence is compensated by any of the remaining terms in Eq.
(\ref{Geod2}) for intermediate values of $a$.

\begin{figure}
\resizebox{\hsize}{!}{\includegraphics{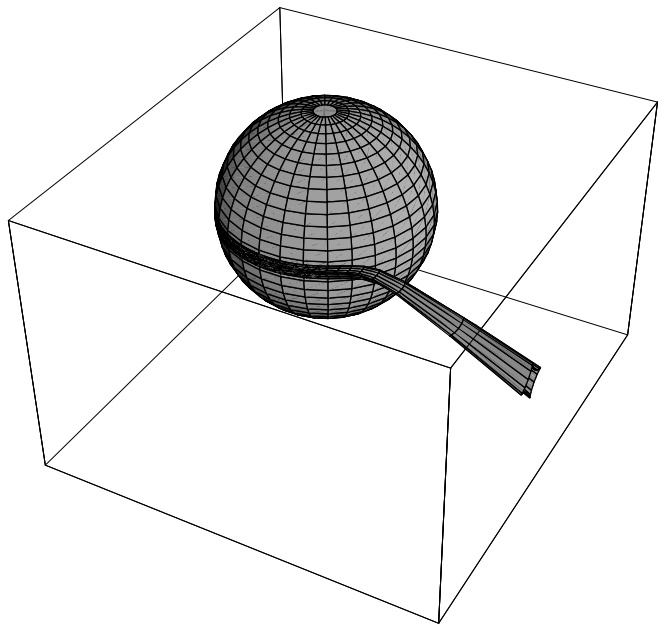}} \caption{A
3-dimensional view (in Boyer-Lindquist coordinates) of the whole
caustic tube for $k=3$, $a=0.1$ and equatorial observer $\mu_O=0$.
The sphere is the horizon of the black hole.}
 \label{Fig Cau3D}
\end{figure}

\begin{figure*}
\resizebox{\hsize}{!}{\includegraphics{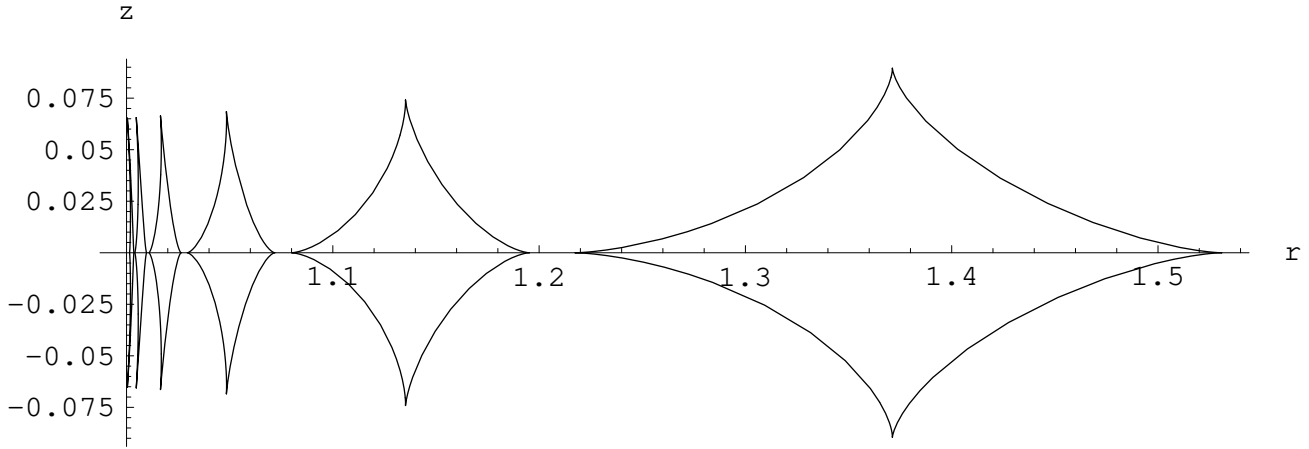}}
\caption{Transverse sections of the caustic tube for $k=3$,
$a=0.1$ and $\mu_O=0$ obtained at
$\delta\phi=0.2,0.3,0.4,0.5,0.6,0.7$ going from the right to the
left. The abscissa is the Boyer-Lindquist radial coordinate,
whereas the ordinate is the pseudo-euclidean coordinate
$z=r\cos\vartheta$.}
 \label{Fig CauSect}
\end{figure*}

Our formulae (\ref{Caumu2}) and (\ref{Caugamma2}) represent the
transverse section of the caustic tube at fixed $D_{LS}$. One
might be interested to a different transverse section, e.g. the
section at fixed $\phi_S$. This is particularly interesting to
study the approach of the caustic tube to the horizon, when the
drift is large and $D_{LS}$ is very close to 1. In this
approximation, we can find the following expression for any fixed
value of $\phi_S$
\begin{eqnarray}
& \mu_S&= (-1)^k\mu_O \pm R_k(D_{LS}=1)\sqrt{1-\mu_O^2}
(1-\xi^2)^{3/2}, \nonumber \\ &&
\label{Caumuphis} \\
& D_{LS}&=1+\frac{16}{9}e^{-\Delta\phi}\left\{1\right. \nonumber
\\ && \left. +\frac{a\sqrt{1-\mu_O^2}}{9}\left[2c_k-(2c_k+7\sqrt{3})\cos^3
\xi\right]\right\} \label{CauDLS},
\end{eqnarray}
where $\Delta\phi=\phi_S-\Delta\phi_k(D_{LS}=\infty)$ is the
difference between the chosen value of $\phi_S$ and the asymptotic
drift of the caustic center. This expression is strictly valid for
large values of $\Delta\phi$, but since the caustic drift sensibly
increases only when $D_{LS}$ is very close to 1, it is sufficient
that $\Delta\phi>0.1$ in order to validate this expansion. Fig.
\ref{Fig CauSect} shows the transverse sections of a caustic tube
obtained at different values of $\Delta\phi$. As the drift is
increased, the caustic becomes thinner and thinner while it
approaches the horizon.

\subsection{Gravitational lensing near caustics} \label{Sec Kerr
lensing}

The higher order images of ordinary sources like stars or X-ray
binaries are usually very faint, except for the event of a caustic
crossing. Therefore, although in principle it is possible to
analyze the lens equation in the general case, it is much more
interesting to study the gravitational lensing of a source in the
neighborhood of a caustic. This case is certainly the most
relevant for observations and is worth a complete and detailed
analysis.

\subsubsection{Position of the images}

The position of a source near a caustic can be expanded in the
following way
\begin{equation}
\mu_S=(-1)^k \mu_o+\delta \mu_S, \label{mu_cau}
\end{equation}
\begin{equation}
\phi_S=(1-k)\pi-\Delta\phi_k+\delta \phi_S , \label{gamma_cau}
\end{equation}
with $\delta\mu_S$ and $\delta\phi_S$ being of the same order of
the caustic extension $R_k$, thus weighing as $a^2$ in the
perturbative expansion.

Correspondingly, the solutions of the lens equation will be very
close to the critical points. We can thus use the following
expansion for $\psi$
\begin{equation}
\psi=k \pi+\psi_{k}^{(1)}+ \delta \psi,
\end{equation}
with $\delta\psi$ being of order $a^2$.

Using these expansions in Eq. (\ref{LensApp}), $\delta \psi$ can
be obtained as a function of the source position and $\xi$
\begin{eqnarray}
&\delta\psi&=-(1-\mu_O^2)\frac{\delta\phi_S}{\hat\xi}-\frac{16}{9}a^2(1-\hat\xi^2)-\frac{c_k}{2}a^2(3-2\mu_O^2-\hat\xi^2)
\nonumber \\
&&+\left[3\sqrt{3}\sqrt{D_{LS}}(2D_{LS}-3)(3+D_{LS})^{3/2}\right]^{-1}
\nonumber \\
&& \cdot \left\{ (3-2\mu_O^2-\hat\xi^2)(3+D_{LS})(2D_{LS}-3)
\right.\nonumber
\\
&& \cdot
\left[9-2D_{LS}-2(2D_{LS}-3)(\sqrt{D_{LS}}-\sqrt{3+D_{LS}})^2\right]
\nonumber
\\
&&\left. -4D_{LS}(3+D_{LS})(2D_{LS}-3)(5-\hat\xi^2)+108D_{LS}
\hat\xi^2 \right\}, \nonumber \\ &&
\end{eqnarray}
whereas $\xi$ is determined by the equation
\begin{equation}
S(-1)^k\frac{\delta\mu_S}{\sqrt{1-\mu_O^2}\sqrt{1-\xi^2}}+\frac{\delta\phi_S\sqrt{1-\mu_O^2}}{\xi}+R_k=0,
\label{xiEq}
\end{equation}
in which $S=\pm 1$ is inherited by the sign ambiguity of the
$(\psi,\xi)$ parametrization (see \cite{KerGen} for more details
about the resolution of the sign ambiguity). Note that the lens
equation formally remains the same as in the $D_{LS}=\infty$ case,
though the position and the extension of the caustic have changed.
The $\xi$ equation (\ref{xiEq}) can be easily put in the form of a
fourth degree polynomial equation. It admits two solutions if the
source is outside the caustic and four solutions if the source is
inside the caustic. The solutions so obtained satisfy the original
equation (\ref{xiEq}) with one definite choice of the sign $S$,
which is then univocally determined for each image.

Once we have the position of the image in the $(\psi,\xi)$ space,
it is straightforward to write the position of the image in the
observer sky. To this purpose, it is important to note that $\xi$
is known through Eq. (\ref{xiEq}) to zero order only. So, the
position of the image on the observer sky is consistently
determined to zero order as
\begin{eqnarray}
&D_{OL} \theta_1=& -\frac{3 \sqrt{3}}{2} \xi(1+\epsilon_{k})
\label{2image1} \\ &D_{OL} \theta_2=& S \frac{3 \sqrt{3}}{2}
\sqrt{1-\xi^2}(1+\epsilon_{k}), \label{2image2}
\end{eqnarray}
with $\epsilon_k$ given by Eq. (\ref{epsk}). The image appears on
the critical ring of order $k$ at a position angle $\arcsin \xi$
and half-sky determined by the sign $S$.

\subsubsection{Brightness of a lensed image}

In typical gravitational lensing studies, the change in the
apparent brightness of the source is simply given by the
geometrical magnification, defined as the ratio of the elementary
angular area of the image and the angular area of the source as it
would be seen without any gravitational lensing.

When the source is far from the lens, this definition makes sense,
since the background metric is asymptotically flat and we can
speak of a source without the lens by simply re-interpreting the
source coordinates as coordinates in the asymptotic Minkowski
metric. This procedure loses any meaning when the source is deeply
within the gravitational field of the black hole. Moreover, the
frequency of the emitted photon does not coincide with the
frequency detected by the observer because of gravitational
redshift. Conservation of the photon number warrants that the
quantity $I/\nu^3$ is conserved on a bundle of light rays, with
$I$ being the specific intensity, defined as the energy $dE$
crossing a surface element $dA$ pointing an angular area $d\Omega$
in the time interval $dt$ and frequency interval $d\nu$.

In order to build a simulated lensed image of a source close to a
black hole, one just needs to find the position of the image for
any point belonging to the source and determine the specific
intensity observed at that point. The position of the higher order
images for any given source position can be read from Eqs.
(\ref{2image1}) and (\ref{2image2}), whereas the specific
intensity measured by the observer is related to the specific
intensity emitted by the source through the relativistic relation
\begin{equation}
I_o=\frac{\nu_o^3}{\nu_e^3}I_e,
\end{equation}
where the redshift factor can be calculated as usual as
\begin{equation}
\frac{\nu_o}{\nu_e}=\frac{p_0}{u^\mu p_\mu},
\end{equation}
with $p_\mu=g_{\mu\nu}\dot x^\nu$ being the momentum of the photon
and $u^\mu$ being the 4-velocity of the emitting particle. In
stationary spherically symmetric and in Boyer-Lindquist
coordinates for the Kerr metric, $\partial_t$ is a Killing vector
and thus $p_0$ is a conserved quantity (we have put it to 1 by a
choice of the affine parameter).

So, for any point of the source we can find the location of the
corresponding high order images and if any model provides us the
specific intensity of the source at that point, we can calculate
the specific intensity as seen by the observer.

In the case in which the source is transparent, it can be
conveniently characterized by its volume emissivity
$j_\nu(x^\mu,\hat p_\mu)$, defined as the energy $dE$ in the
frequency interval $d\nu_e$ emitted by the proper spacetime volume
$\sqrt{-g}d^4 x$ centered on $x^\mu$ in a solid angle $d\Omega$
centered on the direction given by the vector $\hat p_\mu$.

If we want the specific intensity measured by the observer in the
sky direction $(\theta_1,\theta_2)$, we just have to trace back
the null geodesic reaching the observer with such a direction and
sum up the contributions given by all source elements along this
geodesic. We thus have \cite{JarKur}
\begin{equation}
I_o=\int \frac{\nu_o^3}{\nu_e^3} j_\nu(x^\mu,\dot x_\mu)dl_{prop},
\end{equation}
where $dl_{prop}$ is the geodesic line element as measured in the
emitter frame.


\section{Conclusions} \label{Sec Conclusions}

Year after year, our knowledge on the environment surrounding the
supermassive black holes is growing at a higher and higher rate,
thanks to the surprising development of the technology related to
high resolution observations. Angular resolutions of the order of
the microarcsecond are now reachable in the radio band and sooner
or later will be reached in the sub-mm and X-ray band. At the same
time, the interferometric observations in the infrared at the Keck
telescopes and at VLT are unveiling a very rich stellar enviroment
around the Galactic center, which is drawing more and more
interest. High resolution observations are the premise for the
discovery of possible signatures of general relativity effects
from supermassive black holes, which would open a new era in the
understanding of gravitational physics.

Within this context, the comprehension of the propagation of
photons in a strong field environment is of capital importance.
Many numerical codes partially exploiting the analytical solutions
of Kerr geodesics are available and have been used to build
simulated images of the black holes. At the same time, analytical
studies have been complementarily developed to conquer precious
insight about the mathematical structure of the lens mapping in
this very special gravitational lensing framework.

The strong deflection limit allows to study gravitational lensing
in the extreme situation of photons travelling very close to the
unstable circular orbit around the black hole. Such photons give
rise to an infinite sequence of additional images which contribute
to the total flux received by the observer by a non-negligible
amount (see Ref. \cite{BecDon} for an estimate of their
relevance). Therefore, the study of such images is far from being
a mere academic exercise, but acquires a striking importance by
the fact that these images carry invaluable information about the
strong gravitational fields around the black hole. Whereas
complete numerical studies of these images are difficult because
of the extreme  accuracy required to follow photons travelling
around the unstable circular orbit, analytical studies benefit
from the great simplification introduced by the strong deflection
limit. Therefore, higher order images represent a unique window
where we can confront simple analytical results from General
Relativity with observations.

In this paper we have removed the traditional limitation of the
strong deflection limit to sources very far from the black hole.
We have thus analytically explored extreme gravitational lensing
of sources close to a black hole for the first time. There is no
limitation to the validity of our results, which are applicable
even to sources just outside the horizon.

In the spherically symmetric case, we have shown that the same
formulae for the deflection of the photon can be applied both to
sources outside the photon sphere and to sources inside the photon
sphere, whose images appear inside the so-called shadow border. We
have specified our formulae to the Schwarzschild case in order to
test the validity of the strong deflection limit throughout the
range of source distances.

For the Kerr black hole, the only modification comes in the
resolution of the radial integrals in the geodesics equations.
This brings to a modification of the size and shape of the
critical curves. We have obtained a complete analytical
description of the caustic tube from infinite distances up to the
horizon, showing that the caustic tube winds indefinitely around
the black hole because of a logarithmic divergence in the
azimuthal geodesic equation. We have also updated the
gravitational lensing of sources near a caustic.

The formulae contained in this paper can be applied to physically
motivated models of sources around a supermassive black hole,
where they can be used to calculate the shape and the brightness
of the higher order images. We leave this interesting task to
future work, contenting ourselves with the complete analytical
derivation of all relevant formulae here.

\begin{acknowledgments}
The authors acknowledge support for this work by MIUR through PRIN
2006 Prot. 2006023491\_003 and by research fund of the Salerno
University.
\end{acknowledgments}

\appendix

\section{Different perturbative parameters for the Strong
Deflection Limit} \label{AppB}

Recently Iyer and Petters \cite{IyePet} have rewritten Darwin's
formula for the deflection angle of a Schwarzschild black hole in
the strong deflection limit in terms of a new parameter $b'$,
defined as
\begin{equation}
b' \equiv 1-\frac{u_m}{u}.
\end{equation}
Recalling the definition of Darwin's perturbative parameter
$\epsilon$ (\ref{ueps}), we get the simple relation
\begin{equation}
b'=\frac{\epsilon}{1+\epsilon}.
\end{equation}

Of course, if we stop at the lowest order in the expansion, the
deflection angle can be indifferently expressed in the equivalent
forms
\begin{eqnarray}
&\alpha & =-\log \epsilon+\log[216(2-\sqrt{3})^2]-\pi +O(\epsilon)
\\ && =-\log b'+\log[216(2-\sqrt{3})^2]-\pi +O(b').
\end{eqnarray}
Both formulae are correct to lowest order, the differences being
stored in the higher order terms in the perturbative expansions.
Iyer and Petters have found that the higher order discrepancy
between these two formulae and the exact deflection angle is
significantly smaller for the $b'$ formula than for Darwin's one
\cite{IyePet}. Of course, by specifying an invertible function
$f_\lambda(\epsilon)$ such that $f_\lambda(0)=0$ and
$f'_\lambda(0)\neq 0$, one can always define $\lambda \equiv
f_\lambda(\epsilon)$ as a new perturbative parameter. By a
suitable choice of $f_\lambda$ one can make the higher order
corrections vanish up to an arbitrary order $n$. However, this
normally corresponds to a more and more complex form of
$f_\lambda$, which spoils the advantages of the perturbative
expansion. The choice of the perturbative parameter is arbitrary
within the mentioned constraints on $f_\lambda$, but is normally
driven by some physical quantities that can be easily expressed in
terms of the perturbative parameter.

In the case of gravitational lensing in the strong deflection
limit, this point can be made clearer once we construct the
formulae for the position of the images in the two perturbative
frameworks. If we use the $b'$ formula in the lens equation, we
trivially obtain $b'_n=\epsilon_n$ for the $n$-th image, with
$\epsilon_n$ always given by the expression (\ref{epsn}). The
difference between the two frameworks actually emerges from the
expression of the angular position in the observer sky $\theta$ in
terms of the new parameter, which reads
\begin{equation}
\theta=\frac{u}{D_\mathrm{OL}}=\theta_m(1+\epsilon)=\frac{\theta_m}{1-b'}.
\label{thetab'}
\end{equation}
If we plug $b'_n=\epsilon_n$ directly into Eq. (\ref{thetab'}) we
have a new formula for the position of the images
\begin{equation}
\theta_n=\frac{\theta_m}{1-\epsilon_n}, \label{NewTheta}
\end{equation}
which can be compared to the classical formula by Darwin
\begin{equation}
\theta_n=\theta_m(1+\epsilon_n). \label{OldTheta}
\end{equation}
In both formulae $\epsilon_n$ is given by Eq. (\ref{epsn}) and is
just a function of the source position, with no memory of the
perturbative framework used.

If $\epsilon_n$ is small, the two formulae are very close each
other, the difference being order $\epsilon_n^2$ and thus
completely negligible. If an $\epsilon_n^2$ accuracy is necessary,
{\it both} formulae must be complemented by their respective
higher order terms. Then the difference will be in the third order
and so on.

On the other hand, in intermediate situations, in which
$\epsilon_n$ is not small, Eq. (\ref{NewTheta}) does significantly
better than Eq. (\ref{OldTheta}) and can be used to extend the
range of validity of the first order expansion of the Strong
Deflection Limit, without resorting to the second order terms.

\section{Additional terms in the radial integrals} \label{AppA}

As anticipated in Section \ref{Sec GenOut}, the dependence on the
source and observer distances entirely comes from the resolution
of the radial integrals (\ref{I1}) and (\ref{I2}). These integrals
can be performed using the general tools of Section \ref{Sec
Spherical} and then expanded to second order in the black hole
spin $a$. Actually, the integral $I_2$ is already multiplied by
$a$ in Eq. (\ref{Geod2}), so that it is sufficient to stop at the
first order in its expansion. The results are
\begin{eqnarray}
& I_1=&- a_1 \log \delta+ b_1+c_1(D_{LS})+c_1(D_{OL})\\
& I_2=&- a_2 \log \delta+ b_2+c_2(D_{LS})+c_2(D_{OL}),
\end{eqnarray}
with the coefficients $a_1$, $b_1$, $a_2$, and $b_2$ being
unchanged with respect to the appendix of Ref. \cite{KerGen}. To
the second order in $a$, the function $c_1$ reads
\begin{eqnarray}
&c_1(r)&= \frac{a_1}{2} \log \left[(2+\sqrt{3})\frac{\sqrt{3r}-\sqrt{3+r}}{\sqrt{3r}+\sqrt{3+r}} \right]\nonumber \\
&&+ \frac{8a\hat\xi\sqrt{r}}{3\sqrt{3}\sqrt{3+r}(2r-3)} \nonumber
\\
&& +a^2\left[27(2r-3)^2\sqrt{r}(3+r)^{3/2}\right]^{-1} \nonumber
\\ && \cdot \left\{2(2r-3)(3+r)\left[2(2r-3)(\sqrt{r}-\sqrt{3+r})^2(1-\hat\xi^2) \right. \right. \nonumber \\
&& \left. \left. +14r+10r\hat\xi^2-9(1-\hat\xi^2)+8r\mu_o^2
\right] -216r\hat\xi^2\right\}. \label{c1}
\end{eqnarray}
To the first order in $a$, the function $c_2$ is
\begin{eqnarray}
&c_2(r)&=  \frac{a_2}{2} \log \left[(2+\sqrt{3})\frac{\sqrt{3r}-\sqrt{3+r}}{\sqrt{3r}+\sqrt{3+r}} \right] \nonumber \\
&& +\log\left[\frac{2\sqrt{r}+\sqrt{3+r}}{3(2\sqrt{r}-\sqrt{3+r})} \right] \nonumber \\
&&+\frac{2a}{3\sqrt{3}} \hat\xi \left[
\sqrt{r}\sqrt{3+r}(2r-3)\right]^{-1} \nonumber \\
&& \cdot \left[2(2r-3)(\sqrt{r}-\sqrt{3+r})^2+18r-9\right].
\label{c2}
\end{eqnarray}

It can be easily checked that both functions vanish as their
arguments go to infinity. Moreover, if $r\rightarrow 3/2$, the
divergence in $c_i$ is compensated by the vanishing of $\delta$.
As $r<3/2$, it is also easy to see that we must have $\delta^2<0$,
analogously to the spherically symmetric case. The second
logarithm in $c_2$ diverges at $r=1$. The implications of this
divergence are discussed in Section \ref{Sec Kerr Caustics}.

\end{document}